\documentclass[12pt,preprint]{aastex}
\usepackage{graphicx}
\usepackage{epstopdf}
\usepackage{rotating}
\usepackage{bm}

\newcommand{\HI}{\mbox{\sc Hi}}

\begin{document}

\title{Equilibrium Star Formation In A Constant $Q$ Disk: Model Optimisation and Initial Tests}

\author{Zheng Zheng\altaffilmark{1}, Gerhardt R.\ Meurer\altaffilmark{2}, Timothy M. Heckman\altaffilmark{1}, David Thilker \altaffilmark{1}, and 
  Martin Zwaan\altaffilmark{3}}

\altaffiltext{1}{Department of Physics and Astronomy, Johns Hopkins University, 3701 San Martin Drive, Baltimore, MD 21218, USA}
\altaffiltext{2}{International Centre for Radio Astronomy Research, The University of Western Australia, M468, 35 StirlingHighway, Crawley, WA 6009, Australia}
\altaffiltext{3}{European Southern Observatory, Karl-Schwarzschild-Str. 2, 85748 Garching, Germany}

\begin{abstract}
\footnotesize
We develop a  model for  the distribution of the ISM and star formation in galaxies based on recent studies that indicate that galactic disks stabilise to a constant stability parameter, which we combine with prescriptions of how the phases of the ISM are determined and for the Star Formation Law (SFL). The model predicts the gas surface mass density and star formation intensity of a galaxy given its rotation curve, stellar surface mass density and the gas velocity dispersion. This model is tested on radial profiles of neutral and molecular ISM surface mass density and star formation intensity of 12 galaxies selected from the THINGS sample.  Our tests focus on intermediate radii (0.3 to 1 times the optical radius) because there are insufficient data to test the outer disks and the fits are less accurate in detail in the centre. Nevertheless, the model produces reasonable agreement with ISM mass and star formation rate integrated over the central region in all but one case.  To  optimise the model,  we evaluate four recipes for the stability parameter, three recipes for apportioning the ISM into molecular and neutral components, and eight versions of the SFL.  We find no clear-cut best prescription for the two-fluid (gas and stars) stability parameter $Q_{\rm 2f}$ and therefore for simplicity, we use the \citet{wan94} approximation ($Q_{\rm WS}$). We found that an empirical scaling between the molecular to neutral ISM ratio ($R_{\rm mol}$) and the stellar surface mass density proposed by \citet{ler08} works marginally better than the other two prescriptions for this ratio in predicting the ISM profiles, and noticeably better in predicting star formation intensity from the ISM profiles produced by our model with the SFLs we tested. 
 Thus in the context of our modeled ISM profiles, the linear molecular SFL and the two-component SFL \citep{kru09} work better than the other prescriptions we tested. We incorporate these  relations into our `Constant $Q$ disk'  (CQ-disk) model.

\end{abstract}


\section{Introduction}
\label{introduction}

The processes in the interstellar medium (ISM, or gas) that determine its distribution, structure, and the formation of stars are multiple, very complex, and operate on atomic to galactic size scales \citep[see e.g.][]{mck07}.  This makes it difficult to develop a comprehensive model for the distribution of gas and star formation in galaxies from first principles.  Fortunately, normal galaxies are usually in a marginally stable equilibrium state.  Here we use this fact as the basis of a model for the distribution of gas in galaxies, and with some additional assumptions we extend this into a model of the star formation distribution. Our aim is to construct an easy to implement model for the distribution of gas and star formation in galaxies that can be used to compare to observations and to easily create realistic simulated galaxies.   Our approach also allows difficult to observe properties of galaxies (e.g.\ the molecular ISM distribution) to be inferred from those that are relatively easy to determine or infer (the stellar mass profile and rotation curve).

Star formation has long been believed to be related to disk (in)stability.
Major studies \citep[e.g.][]{too64,wan94,raf01,rom11,elm11} have been carried out on gravitational disk stability  since the 1960s and  the theory of this subject is well developed. 
However, as early as 1972, \citeauthor{qui72}  claimed that the whole galactic disk should be marginally stable due to negative feedback mechanism.  He  used this assumption to predict the gas surface density profile; his model roughly matches the observations in the outer regions of his sample galaxies but overestimates the densities in the inner half of the galaxies. 
Since then, observational evidence for Quirk's statement that disk galaxies are usually in a marginally stable state has mounted and been noticed in various studies \citep[e.g.][]{ken89, vdh93, ler08}.

The gravitational stability parameter $Q$ (see section \ref{threshold} for details)  for a
single component thin disk was first derived by
\citet{too64}. \citet{ken89} used a sample of tens of galaxies to show that there is very little star
formation where $Q$ is above some critical threshold. \citet{mar01}
confirmed the result using more recent data. However a single fluid $Q$
does not accurately indicate the stability of a real disk which contains
both stars and gas.  \citet{jog84} derived the two-fluid (stars and gas)
stability parameter $Q_{\rm 2f}$ for a thin disk. \citet{raf01} improved
the derivation by  considering the stellar component as collisionless, 
yielding a rigorous but elaborate form of the stability parameter, $Q_{\rm R}$.  \citet{wan94} proposed an approximation for the two-fluid stability parameter, $Q_{\rm WS}$, which is widely used because of its simple form.  More recently, \citet{rom11} reexamined the \citet{wan94} approximation and gave a simple but more accurate effective two-fluid stability parameters for both an infinitesimally  thin ($Q_{\rm RW,thin}$) and finite thickness  ($Q_{\rm RW,thick}$) galactic disks.

Inspired by the \citet{qui72} and recent studies on disk stability theory and star formation laws, we develop and test the `Constant $Q$ disk' (CQ-disk) model, which we define as: A two-fluid (gas+stars) axisymmetric thin disk will evolve into a marginally stable state with a constant stability parameter $Q_{\rm 2f}$ through out the  galactic disk.  The basic idea for this model is to use the constancy of the two-fluid stability parameter  to predict the distribution of gas and then use empirical molecular-to-neutral gas ratio ($R_{\rm mol}$) relations and the best star formation law (SFL) to predict the distribution of neutral and molecular gas as well as star formation rate. Here we compare the observed distribution of the ISM, and star formation in a small sample of galaxies to our model in order to test our assumptions and to optimise the model by trialling different recipes for its key ingredients ($Q_{\rm 2f}$, $R_{\rm mol}$, and the SFL).

The outline of our paper is as follows: section \ref{bkintro} briefly introduces background information, e.g. the gravitational stability parameter, molecular-to-neutral gas ratio model, and star formation laws; section \ref{model} presents the detailed model we construct;  section \ref{data} gives a brief description of our sample galaxies; section \ref{results} shows the results;  in section \ref{discussion} we discuss the assumption and implications of our model; and in section \ref{application} we discuss further tests and possible uses of our model. 

Through out this paper, we are only concerned with the large scale star formation in the galactic disk, hence we suppose the disk is axisymmetric and focus on the radial profiles of various physical quantities.  
As our research progressed it became useful to divide the galactic disk in to three regions: 
central (within $0.3 r_{25}$),
intermediate ($0.3\sim1 r_{25}$) and outer (  $> r_{25}$), where
$r_{25}$ is the optical radius, defined by where the $B$ band surface
brightness reaches 25 mag arcsec$^{-2}$. 
In general, our model works best in the intermediate radii, and much of the testing of the model is limited to these radii.  The profiles we create often do not match the observations in detail in the central region, although, as we will show, the match to quantities integrated over the central region is reasonable.  
We can not test our star formation models in the outer disk because the star formation intensity was not measured beyond $r_{25}$ in our data sources.  
In \citet{meu13} we discuss the expected ISM structure in the outer disk. There we showed  that a gas dominated constant $Q$ outer disk should have a surface mass  density fall-off with the same profile as the dark matter for systems with a flat rotation curve, thus explaining the close relationship between dark matter and HI \citep{bos81, hoe01, hes11}.
The other reason that we make these divisions is that the RCs are generally more reliable in the intermediate disk: there are less or even no data points in the central regions, while often warps and asymmetries are present  in the outer regions.

\section{Background information}
\label{bkintro}

\subsection{Brief introduction to the gravitational stability parameter}
\label{threshold}
\citet{too64} showed that galactic disks can be unstable to axis-symmetric perturbations, and that the stability of a thin disk to such a perturbation can be represented quantitatively in a single parameter $Q$. Following \citet{too64}, \citet{wan94}, and \citet{raf01}, we express the stability parameter for a gaseous disk:
\begin{equation}
Q_{g}=\frac{\sigma_g \kappa}{\pi G \Sigma_g}, 
\label{Qg}
\end{equation}
where $\sigma_g$ is the gas velocity dispersion. Typically $\sigma_g$ is assumed to be constant \citep[e.g.][hereafter L08]{ler08}, although observations show that it typically slowly varies with radius \citep[e.g.][] {tam09,obr10}.
 $G$ is the gravitational constant, $\Sigma_g$ is the gas surface mass density and $\kappa$ is the epicyclic frequency which can be calculated from the rotation curve using
\begin{equation}
\kappa=\frac{v}{r}\sqrt{2\left(1+\frac{r}{v} \frac{dv}{dr}\right)},
\label{e:kappa}
\end{equation}
where $v$ is the circular velocity of the gas at a distance $r$ from the galactic centre.  Toomre's $Q$  parameter can be an indicator for 
widespread star formation: a large stability parameter ($Q_g>1$) means that pressure and centrifugal forces are sufficient to support the disk and thus it is stable, while a small $Q_g(<1)$ means the gravity exceeds the internal support and the disk will collapse, resulting in widespread star formation \citep{ken89,vdh93,mar01}.

For a single component stellar disk, we can  use a similar  parameter, $Q_s$, to indicate gravitational stability of the stars \footnote{This is originally defined as $Q_{s}=\sigma_{s,r} \kappa/3.36 G \Sigma_s$ \citep{too64,bin08}, but redefined as the equation shown above by later authors e.g.\ \citet{wan94,raf01}  in order to use a more consistent expression with regards to $Q_g$.} : 
\begin{equation}
Q_{s}=\frac{\sigma_{s,r} \kappa}{\pi G \Sigma_s}, 
\label{Qs}
\end{equation}
where $\Sigma_s$ is the star surface mass density and $\sigma_{s,r}$ is the radial component of the stellar velocity dispersion \citep{jog84,wan94,raf01}. We follow the estimation of $\sigma_{s,r}$ given by L08: assuming a typical fixed shape to the velocity dispersion ellipsoid, we have
\begin{equation}
\sigma_{s,r}=1.67\sigma_{s,z}, 
\end{equation}
where $\sigma_{s,z}$ is the z-direction component of the stellar velocity dispersion and using the standard relationship between this, disk scale height $h_s$, and mass density for an isothermal disk \citep[e.g.][]{vdk88}, we have
\begin{equation}
\label{sigma_sz}
\sigma_{s,z}=\sqrt{2\pi G \Sigma_s h_s}.
\end{equation}
Here $h_s$ is assumed to be
a constant through out the galaxy disk (e.g. as done by L08). Since,
$h_s$ is difficult to measure unless the galaxy is viewed edge-on, we follow L08 and use the exponential stellar disk's scale length, $l_s$, to estimate it. Using the average flattening ratio $l_s/h_s=7.3\pm2.2$, measured by \citet{kre02}, we then obtain 
\begin{equation}
\sigma_{s,r}\approx 1.55 \sqrt{G \Sigma_s l_s}.
\end{equation}

Galactic disks are composed of both gas and stars. 
Thus, we need to consider both of them in a more sophisticated disk stability theory.
There have been numerous studies on the two-fluid disk stability theory and therefore
different forms of two-fluid stability parameters, $  Q_{2f}$. 
Here we introduce four of them:
 
\citet{jog84} explored the gravitational stability of a two-fluid thin disk. Later-on, \citet{raf01} extended this work by treating the stellar part of the disk as collisionless and gave an explicit expression for the total stability parameter $Q_{\rm R}$:
 \begin{equation}
\frac{1}{Q_{R}}=\frac{1}{Q_s}\frac{2}{1/q+q}+\frac{1}{Q_g}\frac{2}{yq+1/yq},
\label{Qr}
\end{equation}
where 
\begin{equation}
y=\sigma_g/\sigma_{s,r},
\label{eq_y}
\end{equation}
 and $q=k\sigma_{s,r}/\kappa$, with $k$ the wave number of the instability being considered. We take $k$ to be the wavenumber of the most unstable mode in our calculations (L08).  In this paper, we make a coarse searching through the parameter space of $k$ and find  the value which gives the lowest gas surface mass density. We find that the galaxies studied here have $\lambda=2\pi/k = 2-5 kpc$, consistent with the result of L08. 

Although the \citet{raf01} derivation of the two-fluid stability parameter is quite rigorous, it is  complicated and  wavenumber dependent, making it hard to use.   \citet{wan94} give a much simpler form, 
$Q_{\rm WS}$:
\begin{equation}
\frac{1}{Q_{\rm WS}}=\frac{1}{Q_g}+\frac{1}{Q_s}.
\label{Qws}
\end{equation}
It is widely used by astronomers \citep[e.g.][]{mar01} although the analysis deriving it has been criticised \citep{jog96}. \citet{rom11}  improved the \citet{wan94} approximation and proposed an effective $Q$ parameter for a two-fluid infinitesimally thin disk, 
$Q_{\rm RW,thin}$:
\begin{equation}
\label{rwthin}
\frac{1}{Q_{\rm RW,thin}}=
\left\{\begin{array}{ll}
       {\displaystyle\frac{W}{Q_{s}}+\frac{1}{Q_{g}}}
                       & \mbox{if\ }Q_{s}\geq Q_{g}\,, \\
                       &                                            \\
       {\displaystyle\frac{1}{Q_{s}}+\frac{W}{Q_{g}}}
                       & \mbox{if\ }Q_{g}\geq Q_{s}\,;
       \end{array}
\right.
\end{equation}
where 
\begin{equation}
W(y)=\frac{2}{y+1/y},
\end{equation}
and $y$ is defined above in Eq. \ref{eq_y}.

In reality, disks are not infinitesimally thin, and the thickness of the disk has a  stabilising effect.
Therefore, \citet{rom11} also give an expression for a disk of finite thickness, 
 $Q_{\rm RW,thick}$:
\begin{equation}
\label{rwthick}
\frac{1}{Q_{\rm RW,thick}}=
\left\{\begin{array}{ll}
       {\displaystyle\frac{W}{T_{s}Q_{s}}+
                     \frac{1}{T_{g}Q_{g}}}
                       & \mbox{if\ \ }T_{s}Q_s \geq
                                      T_{g}Q_{g}\,, \\
                       &                                              \\
       {\displaystyle\frac{1}{T_sQ_s}+
                     \frac{W}{T_{g}Q_{g}}}
                       & \mbox{if\ \ }T_{g}Q_g \geq
                                      T_sQ_s\,;
       \end{array}
\right.
\label{Qrw2}
\end{equation}
where $T$ is a factor, by which the stability parameter of each component increases, reflecting the effect of the thickness of the disk. The factor $T$ depends on the ratio of vertical to radial velocity dispersion: $T= 0.8+0.7(\sigma_z / \sigma_r)$. So that we have $T_g\approx 1.5$ for gas, and $T_s\approx1.22$ for stars.

\citet{ken89} showed that $Q_g$ values for 15 disk galaxies are almost a constant. Similarly \citet{meu13} used more recent measurements to show that $Q_g$ is constant over a large fraction of the outer disks of galaxies where HI  dominates the mass.   \citet{boi03, mar01} tested the \citet{wan94} $Q_{\rm WS}$ on more disk galaxies and got a similar result.  L08 used the \citet{raf01} $Q_{\rm R}$ in testing the star formation threshold for 23 nearby galaxies and get a remarkably flat $Q_{\rm R}$ thorough the optically bright  part of galactic disks with $Q_{\rm R} \sim 1.3-2.5$.  

One plausible explanation for this phenomena is self-regulation by star
formation \citep{bur92, elm11}: an unstable disk (low $Q$) will collapse
and commence widespread star formation. 
The star formation will consume most of the gas in a short time \citep{qui72}, 
also the resulting winds from high mass stars and supernovae explosions 
will expel part of the gas out of the galactic disk \citep{dut09a,hec90} 
and heat up the remaining gas \citep{tam09}. This will increase  $Q$ by lowering $\Sigma_g$ and increasing $\sigma_g$ and thus make the disk more stable. On the other hand, a stable disk (high $Q$) will have 
suppressed star formation because the gas can not collapse.  This means less heating of the ISM through winds and supernovae, and hence a relative cooling and lowering of $\sigma_g$, thus decreasing $Q$. 

Thus a negative feedback system is set up and it will finally come to a stable equilibrium state. Here we suppose that galaxies equilibrate to the same $Q$ at all radii through this local feedback cycle: regions of low $Q$ heat up locally due to feedback, while regions of high $Q$ cool due to relative lack of star formation. We test whether $Q$ is  constant within the optically bright portion of galaxies, and consider whether $Q$ may be constant within a galaxy, but vary from galaxy to galaxy. This might happen if the balance in the local support between epicyclic frequency $\kappa$ and velocity dispersion $\sigma$ varies due to changes in the relative importance of angular momentum and feedback, perhaps due to  initial mass function  variations \citep{meu09} or metallicity dependent cooling.

\subsection{Brief introduction to $R_{\rm mol}$ relations}
\label{rmol}
It is useful to have an accurate prescription for determining the molecular to neutral ratio ($R_{\rm mol}\equiv \Sigma_{\rm H_2}/\Sigma_{\rm HI}$) because these phases of the ISM are measured with separate observations and because it is generally believed that star formation is more related to the the molecular rather than the atomic gas.  In practice the molecular data has been more expensive to obtain than \HI\ data and requires use of tracers of H$_2$ like CO emission.  The advent of new facilities like SMA, CARMA and ALMA may change these economics.

Various simple prescriptions for determining $R_{\rm mol}$ have been raised in previous studies
 (cf. L08 and references therein).  L08 used high quality HI and CO data 
to assess four of the most popular $R_{\rm mol}$ prescriptions and found two that worked well.
The first is purely empirical and  involves just the stellar surface mass density (we call this the SR relation or $R_{\rm mol,s}$)
\begin{equation}
R_{\rm mol,s}=\frac{\Sigma_s}{81\, M_{\odot}\, pc^{-2}}.
\end{equation}
The other uses hydrostatic pressure $P_h$ as a parameter  (the PR relation or $R_{\rm mol,p}$) and is
\begin{equation}
R_{\rm mol,p}= \left(\frac{P_h}{1.7\times10^4\, cm^{-3}\,K\, k_B} \right)^{0.8},
\end{equation}
where $P_h$ is given by \citet{elm89} as
\begin{equation}
P_h= \frac{\pi}{2}G \Sigma_g(\Sigma_g+\frac{\sigma_g}{\sigma_{s,z}}\Sigma_s),
\end{equation}
where $\sigma_{s,z}$ can be calculated using Eq. \ref{sigma_sz}. Although L08 found that $P_h$ is not as good of a predictor of $R_{\rm mol}$ as the stellar surface mass density, it has a more solid physical basis \citep{won02,bli04,bli06,elm89,elm94}. 

Another  physically intuitive model is described by \citet[hereafter KR relation or $R_{\rm mol,K}$]{kru08, kru09a, kru09} and tested by \citet{kru11} using numerical simulations. In this model the molecular gas can only exist in a region well shielded from the UV radiation field. The approximate solution is 
\begin{equation}
R_{\rm mol,K}=\frac{4-2s}{3s},
\label{eq_molk}
\end{equation} 
\citep{kru11}, where 
\begin{equation}
s=\frac{ln(1+0.6\chi+0.01\chi^2)}{0.6 \tau_c},
\label{eq_s}
\end{equation}
and where $\chi$ is a dimensionless number representing the scaled radiation field, 
\begin{equation}
\tau_c=\Sigma_g a_d/\mu_H,  
\end{equation}
where $a_d$ is the dust cross section per hydrogen atom and $\mu_H=2.3\times10^{-24}g$ is the mean mass per nucleus. Following \citet{kru11}, we take the approximation that 
\begin{equation}
\frac{\sigma_{d}}{10^{-21} cm^{-2}}=Z', 
\label{eq_crosssection}
\end{equation}
and 
\begin{equation}
\chi \approx 3.1 \left( \frac{1+3.1 Z'^{0.365}}{4.1} \right),
\label{eq_zprime}
\end{equation}
with $Z'$ being the metallicity normalised to the solar value. From Eq. \ref{eq_molk} - \ref{eq_zprime} we expect some troublesome asymptotic behaviour in low density and low metallicity regions because a small $\tau_c$ ($\ll 0.5$) leads to a negative $R_{\rm mol,K}$. We therefore set negative $R_{\rm mol,K}$  values to zero in our code, which means there is no molecular gas in low density and low metallicity regions.

We test all three relations to estimate $R_{\rm mol}$ in our calculation ( using the total gas surface mass density derived using our model, cf model description in section \ref{model} ) and see which prescription best models the $\Sigma_{\rm HI}$ and $\Sigma_{\rm H_2}$ profiles.

\subsection{Brief introduction to star formation laws}
The SFL is usually defined as the relationship between star formation rate and ISM properties, especially density. The SFL allows one to calculate the star formation rate given gas density and some other galactic parameters or vice versa. 
This is usually expressed as the surface density of star formation, $\Sigma_{\rm SFR}$, defined as the SFR per unit area, or in other words the star formation intensity.
There have been numerous SFLs and here we summarise and test the most popular ones in the literature \citep[L08;][and references therein]{big08,tan10}.

\label{SFLs}
\subsubsection{Schmidt-Kennicutt Law}
Probably the most well known and widely used SFL is the Schmidt-Kennicutt Law \citep{sch59, ken98}. It is a simple empirical correlation between star formation intensity and total gas:
\begin{equation}
\Sigma_{\rm SFR,SK}=A_{\rm SK} \left (\frac{\Sigma_g}{\rm 100 M_{\odot}\, pc^{-2}}\right )^N,
\end{equation}
where $A_{\rm SK}$ is a coefficient with unit of [${\rm M_{\odot}\, kpc^{-2}\, yr^{-1}}$],  $\Sigma_g$ is the surface mass density of the gas (both neutral and molecular), 
and $N=1.4\pm0.15$ \citep{ken98}. Here we fix $N = 1.4$ in our tests of $\Sigma_{\rm SFR,SK}$.

\subsubsection{Free-fall time scale with fixed scale height}
The power law index $N$ of Schimidt-Kennicutt Law is very close to 1.5, which can be explained by arguing that stars form in a free-fall time scale in a gas disk with fixed scale height: Since the free fall time scale $\tau_{\rm ff}$ is proportional to the inverse square root of the local gas density $\rho_g$, then for a fixed scale height $h$, the star formation intensity is \citep{ler08}:
\begin{equation}
\Sigma_{\rm SFR}\propto \frac{\Sigma_g}{\tau_{\rm ff}} \propto \frac{\rho}{\rho^{-0.5}} \propto\Sigma_g^{1.5}.
\end{equation}
Therefore the star formation intensity in a free-fall timescale with fixed scale height can be written as
\begin{equation}
\Sigma_{\rm SFR,ff}=A_{\rm ff}  \left (\frac{\Sigma_g}{\rm 100 M_{\odot}\, pc^{-2}}\right )^{1.5},
\end{equation}
where $A_{\rm ff}$ has the same unit as $A_{\rm SK}$. So this version of the SFL is nearly the same as the Schmidt-Kennicutt SFL with a slightly higher $N = 1.5$.

\subsubsection{Free-fall timescale with variable scale height}
If the disk scale height is not fixed but set by hydrostatic equilibrium, then the star formation law (still assuming stars form in disk free-fall time scale) is \citep{ler08}
\begin{equation}
\Sigma_{\rm SFR,ff'}=A_{\rm ff}^{\prime} \left (\frac{\Sigma_g}{\rm 100 M_{\odot}\, pc^{-2}} \right)^2 \left (1+\frac{\Sigma_s}{\Sigma_g} \frac{\sigma_g}{\sigma_{s,z}}\right )^{0.5}  \left (\frac{11km/s}{\sigma_g} \right),
\end{equation}
where $A_{\rm ff}^{\prime} $ has the same unit as $A_{\rm SK}$.

\subsubsection{Orbital timescale}
If we instead assume that a constant fraction of the ISM is consumed in a dynamical timescale, 
we would have \begin{equation}
\Sigma_{\rm SFR,\Omega}=B_{\Omega}\Sigma_g \Omega,
\end{equation} 
where $B_{\Omega}$ is a dimensionless coefficient and $\Omega$ is the orbital angular frequency \citep{ler08, tan10}. Following the suggestion of this functional form by \citet{silk97} and \citet{elm97}, \citet{ken98} showed that this form of the SFL was able to account for the star formation intensity in his sample of normal galaxies and circum-nuclear star burst, and worked equally well as $\Sigma_{\rm SFR,SK}$.

\subsubsection{GMC collisions in a shearing disk}
\label{sfl_cc}
\citet{tan00} presented a star formation model which assumes that star formation is triggered by GMC collisions in a shearing disk. This predicts that
\begin{equation}
\label{sfrcoll}
\Sigma_{\rm SFR,CC} = B_{\rm CC} Q_g^{-1}  \Omega (1 - 0.7 \beta) \Sigma_g \:\:\: (\beta \ll 1),
\end{equation}
where  $B_{\rm CC}$ is a dimensionless coefficient, $\beta\equiv d\:{\rm ln}\:v_{\rm circ} / d\:{\rm ln}\: r$ and 
$v_{\rm circ}$ is the circular velocity at a particular galactocentric
radius $r$. Note $\beta=0$ for a flat rotation curve. 
The GMC collision SFL is only meant to be valid in regions where a significant fraction of the gas is in gravitationally bound clouds, which typically means $  \Sigma_{\rm H_2} \gtrsim   \Sigma_{\rm HI}$. However, here we test it over a wider range of densities.

\subsubsection{Linear molecular SFL}
\label{sfl_h2}
\citet{ler08} and \citet{big08} studied 12 nearby spiral galaxies at sub-kpc resolution and concluded that the SFR is proportional to the molecular content: 
\begin{equation}
\Sigma_{\rm SFR,H2}=A_{\rm H_2}\left(\frac{\Sigma_{\rm H_2}}{\rm 100 M_{\odot}\, pc^{-2}}\right),
\end{equation}
where the coefficient $A_{\rm H_2}$ has the same unit as $A_{\rm SK}$.


\subsubsection{Turbulence-regulated SFL}
\label{sfl_turbulence}
\citet{kru05} provided a  turbulence-regulated star formation model to predict the SFR by assuming stars primarily form in molecular clouds that are virialised and supersonically turbulent and that the probability distribution of densities is lognormal. The intensity of star formation by their model is given by
\begin{equation}
\Sigma_{\rm SFR,KM}= A_{\rm KM} f_{\rm GMC} (\frac{\phi_{\bar{P}}}{6}) ^{0.34} (\frac{Q_g}{1.5}) ^{-1.32} (\frac{\Omega}{\rm Myr^{-1}})^{1.32} (\frac{\Sigma_{g}}{\rm 100 M_{\odot}\, pc^{-2}}) ^{0.68},
\end{equation}
where the coefficient $A_{\rm KM}$ has the same unit as $A_{\rm SK}$, 
$f_{\rm GMC}$ is the mass fraction of gas in giant molecular clouds (GMCs) and can be approximated as $f_{\rm GMC} =R_{\rm mol}/(1+R_{\rm mol})$, and $\phi_{\bar{P}}=10-8f_{\rm GMC}$ \citep{kru05, tan10}.

\subsubsection{Two-component SFL}
\label{sfl_2comp}
\citet{kru09} presented a two-component star formation law, 
\begin{eqnarray}
\label{sfrKMT}
\Sigma_{\rm SFR,KMT}  =  A_{\rm KMT} f_{\rm GMC} \left( \frac{\Sigma_g}{\rm 100 M_{\odot}\, pc^{-2}} \right) \times \left\{ 
\begin{array}{lc}
\left(\Sigma_g/\Sigma_0\right)^{-0.33}, & \Sigma_g< \Sigma_0\\
\left(\Sigma_g/\Sigma_0\right)^{0.33}, & \Sigma_g> \Sigma_0
\end{array}
\right\},
\end{eqnarray}
where the coefficient $A_{\rm KMT}$ has the same unit as $A_{\rm SK}$,
and $\Sigma_0=85\:{\rm M_\odot pc^{-2}}$ is a `critical' gas surface mass density. In regions with $\Sigma_g < \Sigma_0$, GMCs have an internal pressure that far exceeds the ambient gas pressure and the star formation time scale is independent of the environment; whilst in regions with $\Sigma_g > \Sigma_0$,  the star formation time scale depends  on the metallicity and the clumping.

\section{The Model}
\label{model}

Our model requires at least three inputs: the rotation curve, gas velocity dispersion and the stellar mass distribution. The metallicity is also needed if using the KR relation (cf. \ref{rmol}).  
 The basic assumption for our model is that the  galactic disk is in a gravitationally marginal stable state and the $R_{\rm mol}$ relations and SFLs are valid for all the galaxies.

Our algorithm is as follows:

1). Determine the radial total gas distribution  ($\Sigma_g(r)$) assuming a marginally stable disk.

We make the assumption that the two-fluid stability parameter 
, $  Q_{\rm 2f}$ , 
is constant through out the whole galactic disk. We test all  four two-fluid stability parameters ($Q_{\rm R}$, $Q_{\rm WS}$, $Q_{\rm RW,thin}$, $Q_{\rm RW,thick}$, 
 cf section \ref{threshold}
) on  our sample galaxies and pick out the one most suitable for our model and then adopt it for all further calculations. We define `most suitable' in the first place as having the flattest Q profile with the smallest rms deviation over all or most of the sample galaxies, and as a secondary consideration the simplest form.

Based on eq. (\ref{Qg})-(\ref{Qrw2}), we  calculate the gas surface mass density $\Sigma_g$ from the rotation curve $v(r)$, gas velocity dispersion and stellar surface mass density $\Sigma_s$ by fixing the stability parameter $Q_{\rm 2f}$ to the average value at intermediate radii.

2). Determine the radial molecular ($\Sigma_{\rm H_2}(r)$) and  neutral ($\Sigma_{\rm HI}(r)$)  gas distribution  

We test three different ways to determine $R_{\rm mol}$: the SR, PR and KR relations (cf. section \ref{rmol}).
Once we have $R_{\rm mol}$ we  determine the molecular gas surface mass density 
 based on our model predicted $  \Sigma_{\rm g}$ from step 1
\begin{equation}
\Sigma_{\rm H_2}=\Sigma_g \frac{R_{\rm mol}}{1+R_{\rm mol}},
\end{equation}
and neutral gas surface mass density 
\begin{equation}
\Sigma_{\rm HI}=\Sigma_g \frac{1}{1+R_{\rm mol}}.
\end{equation}
We then compare the derived molecular and neutral gas surface mass density  to the observations to determine which $R_{\rm mol}$ relation fits the data best.

3). Determine the star formation intensity ($\Sigma_{\rm SFR}(r)$)

After obtaining the $\Sigma_{\rm HI}$ and $\Sigma_{\rm H_2}$, we use the SFL(s) introduced in section \ref{SFLs} to calculate the star formation intensity $\Sigma_{\rm SFR}$.  Following \citet{tan10}, we fit each galaxy using every SFL described in section \ref{SFLs}.  The results are evaluated to determine which SFL(s) fit the observed $\Sigma_{\rm SFR}$ best so as to implement in future application of our model.

\section{The Data}
\label{data}
In order to construct our model, we need the following  quantities as inputs: the rotation curve $v(r)$, the stellar surface mass density $\Sigma_s$, the gas velocity dispersion $\sigma_g$, and metallicity. Also, for comparison with the observations, we need data for the neutral gas surface mass density $\Sigma_{\rm HI}$, the molecular gas surface mass density $\Sigma_{\rm H_2}$ and the star formation intensity $\Sigma_{\rm SFR}$.

The best sample, which has all the data listed above, is a subset of the THINGS sample (L08; de Blok et al. 2008). It is composed of five dwarf galaxies (DDO 154, IC 2574, NGC 7793, NGC 2403, and NGC 925) and seven spiral galaxies (NGC 3198, NGC 4736, NGC 6946, NGC 3521, NGC 5055, NGC 2841, and NGC 7331). The $\Sigma_s$, $\Sigma_{\rm HI}$ and $\Sigma_{\rm H_2}$ are tabulated in the paper of L08, although the dwarf galaxies do not have CO data and thus do not have $\Sigma_{\rm H_2}$. The $v(r)$ is provided by \citet{deb08}.

L08 used a fixed value of $\sigma_g (=11 km/s)$. However, 
in general $\sigma_g$ is observed to vary in galaxies  \citep{tam09,obr10}. 
Therefore, we tried both the measured $\sigma_g$ and a fixed $\sigma_g=11\, {\rm km/s}$ when calculating the $Q_{\rm 2f}$.
The measured $\sigma_g$ are from A. C. Primo \& F. Walter (private communication) and plotted in Fig. \ref{fig_vsig}

The metallicity data is taken from \citet{mou10}. We use a linear fit for each galaxy based on its metallicity zero point and radial gradient  \citep{kob04} provided by Moustakas (private communication). The solar metallicity is taken as $12+\log$(O/H)=$8.69$ \citep{mou10,asp09}.

For the rotation curve $v(r)$, we fit the data using the universal rotation curve (URC) \citep{bat00,per96}. The URC can be parameterised as 
\begin{equation}
v^2(r)=v_0^2\beta \frac{1.97x^{1.22}}{(x^2+0.78^2)^{1.43}}+v_0^2(1-\beta)(1+a^2)\frac{x^2}{x^2+a^2},
\end{equation}
where $x=r/r_{opt}$ is the radial variable. Nominally, $r_{opt}$ is the radius encircling $83\%$ of the light, $v_0$ is the velocity at $r=r_{opt}$, while $\beta$ and $a$ are constants that depend on the luminosity. Here we keep the $v_0$, $r_{opt}$,  $\beta$, and $a$ as free parameters.

\begin{figure}
\begin{center}
\includegraphics[scale=0.8]{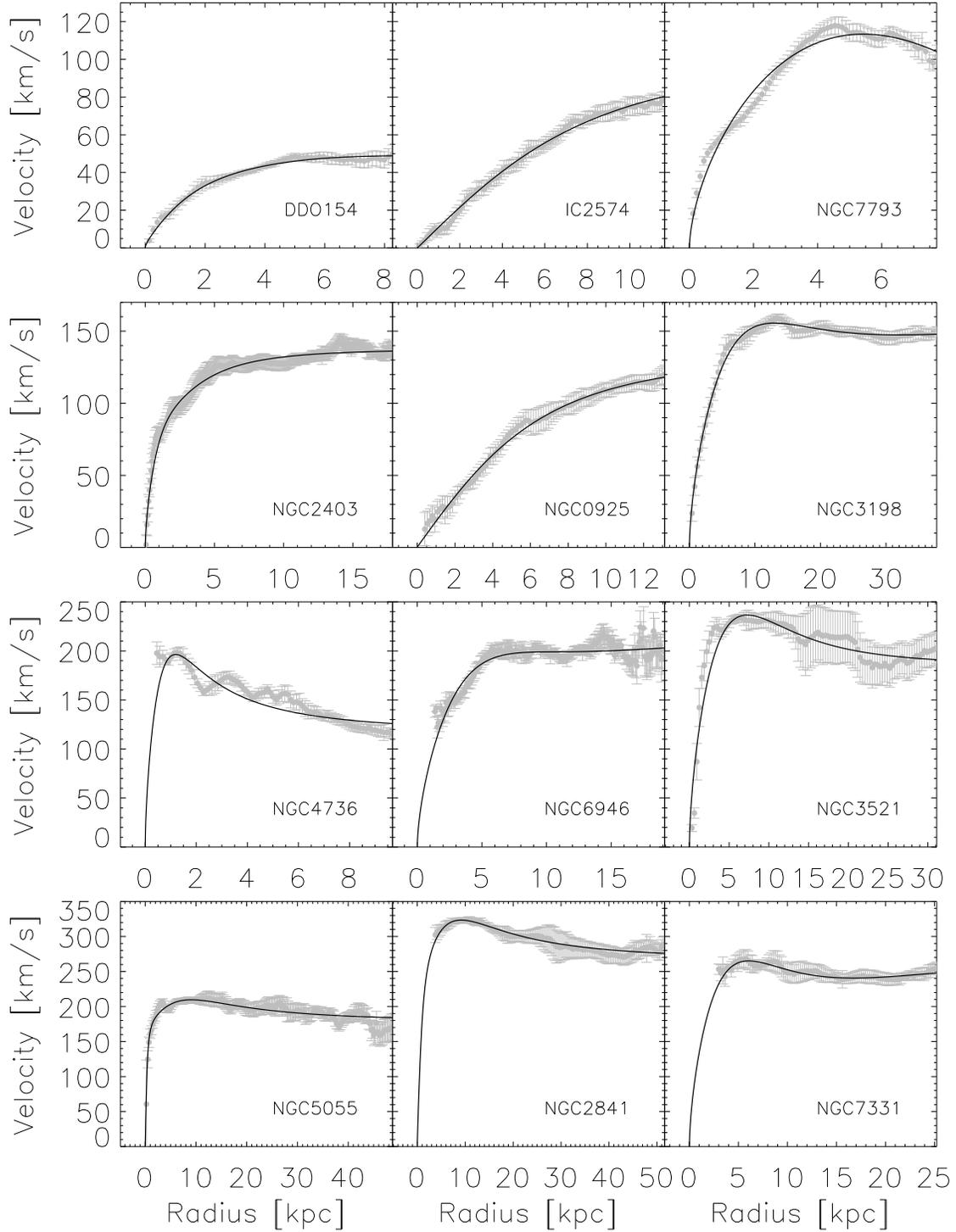}  \caption{Rotation curve data and URC fitting. The grey dots with error bars are measured rotation curve data with uncertainty and the black solid lines are the URC fits. }
\label{vrot}
\end{center}
\end{figure}

The rotation curve data and their URC fits are shown in Fig. \ref{vrot}. The main advantage of using such a smooth rotation curve fit is to have a well defined and realistic derivative of $v(r)$. This is important when calculating $\kappa$ (eq.~\ref{e:kappa}) and thus essential for any version of $Q_{\rm 2f}$.  We note that an overly tight fit to all the kinks and wiggles in the rotation curve amplifies these features in the derivative, producing unrealistic results, especially when the fitted $v(r)$ profile declines faster than Keplerian.  It is likely that these small amplitude variations in $v(r)$ result from noise in the data and small scale non-circular motions due to bars, spiral arms, and asymmetries.

The URC parameters are fitted using  the {\tt MPFIT} software
package \citep{mar09} to $v(r)$.
We also tried two other rotation curve parameterisations: the exponential form (L08) and  the arctan form \citep{cou97}.  Generally, the URC model fits the data best, 
therefore we use the URC fit for all the rotation curves in this paper. 
The URC model fitting sometimes results in unphysical
parameters (e.g. the fitted $R_{\rm opt}$ of DDO 154 is 12.1 kpc,  far greater than the observation, 1.2 kpc; L08), but we use them anyway
because we are only concerned with getting a smooth form which can best
represent the data. 

\section{Results}
\label{results}

\subsection{The two-fluid stability parameter $Q_{\rm 2f}$}

The four two-fluid Q  models  are fitted to a constant (in forms of $\log Q$) using the {\tt MPFIT} software package.  
In order to assess the quality of the fit we use the ``deviation value'' $\epsilon_Q$ which we define as 
\begin{equation}
\label{eq_epsilon_Q}
\epsilon_Q= \sqrt{ \frac{1}{N_{\rm ann}-1} \sum_{i=1}^{N_{\rm ann}} \left(\log \frac{Q_{\rm model}}{Q_{\rm obs}}\right) ^2       }  ,                    
\end{equation}
where $N_{\rm ann}$ is the number of valid annuli in the galaxy. $\epsilon_Q$ has units of dex; values of $\epsilon_Q$ closest to zero indicate the model that most closely represents the observations.  The fitted values and their corresponding deviation values are shown in Table~\ref{qerr}.  Because the observed $Q_{\rm 2f}$ are less constant in the central and outer regions, we only use the data within the intermediate disk during the fitting.  In terms of flatness, the average $\epsilon_Q$ listed at the bottom of the Table~\ref{qerr} indicate that $Q_{\rm R}$ is marginally the best recipe, followed by $Q_{\rm RW,thick}$, $Q_{\rm WS}$, and $Q_{\rm RW,thin}$. It is satisfying that the form of $Q_{\rm 2f}$ that is most rigorously defined yields the lowest average $\epsilon_Q$, because it suggests that the feedback processes are truly working to maintain disks at a constant stability, at least over the intermediate radii.  But since all these $Q_{\rm 2f}$ recipes result in typically a factor of $\sim 1.6$ deviations about the mean, we choose to use $Q_{\rm WS}$ in our calculations as the easiest to implement, and sufficient for our purposes.

\begin{deluxetable}{l c c c c c c c c c c c c c c}
  \tabletypesize{\tiny} 
  \tablewidth{0pt} \tablecolumns{14}
  \tablecaption{$Q_{\rm 2f}$ model fit results and galaxy properties}
  \tablehead{\colhead{Galaxy}    & \colhead{$N_{\rm ann}$} &
  \colhead{$Q_{R}$} & \colhead{$\epsilon_{Q_{\rm R}}$} & \colhead{$Q_{\rm WS}$} & \colhead{ $\epsilon_{Q_{\rm WS}}$} 
  & \colhead{  $Q_{\rm RW,thin}$}   & \colhead{  $\epsilon_{Q_{\rm RW,thin}}$}   & \colhead{  $Q_{\rm RW,thick}$} & \colhead{  $\epsilon_{Q_{\rm RW,thick}}$}  
  & $V_m $  & $\log(M_S) $  & $\log(M_g) $  & $\rm \frac {SFR_{cen}}{SFR_{int}}$ \\
&   & & & & & & & & & [km/s] & $ [\log M_{\odot}]$ & $ [\log M_{\odot}]$  \\
    }
\startdata
    \label{qerr}
DDO154 & 4 &  3.50 &   0.10 &   2.80 &   0.10 &   3.26 &   0.16 &   4.48 &   0.19 &  50.00 &   7.10 &   8.70  & 0.20\\
IC2574 &  27 & 1.46 &   0.21 &   1.26 &   0.20 &   1.30 &   0.21 &   1.84 &   0.21 & 134.00 &   8.70 &   9.30  & 0.13\\
NGC7793 &  22 & 1.77 &   0.19 &   1.70 &   0.21 &   1.79 &   0.18 &   2.34 &   0.19 & 115.00 &   9.50 &   9.10 & 0.55\\
NGC2403 &  33 & 1.96 &   0.18 &   1.89 &   0.20 &   1.98 &   0.17 &   2.61 &   0.17 & 134.00 &   9.70 &   9.50 & 0.49\\
NGC0925 &  22 &  1.82 &   0.22 &   1.36 &   0.23 &   1.44 &   0.27 &   1.98 &   0.25 & 136.00 &   9.90 &   9.82 & 0.43\\
NGC3198 &  14 &  1.82 &   0.16 &   1.42 &   0.15 &   1.58 &   0.23 &   2.09 &   0.21 & 150.00 &  10.10 &  10.12 & 0.37\\
NGC4736 &  16 &  1.85 &   0.27 &   1.73 &   0.29 &   1.78 &   0.28 &   2.20 &   0.29 & 156.00 &  10.30 &   8.95  & 3.45\\
NGC6946 &  24 &  1.55 &   0.26 &   1.21 &   0.31 &   1.53 &   0.23 &   2.05 &   0.20 & 186.00 &  10.50 &  10.01 & 0.31\\
NGC3521 &  18 &  1.57 &   0.17 &   1.19 &   0.21 &   1.50 &   0.20 &   1.99 &   0.19 & 227.00 &  10.70 &  10.15 & 0.66\\
NGC5055 &  25 &  1.67 &   0.18 &   1.31 &   0.28 &   1.69 &   0.20 &   2.29 &   0.16 & 192.00 &  10.80 &  10.25 & 0.83\\
NGC2841 &  15 &  3.99 &   0.17 &   2.52 &   0.19 &   2.94 &   0.21 &   3.66 &   0.20 & 302.00 &  10.80 &  10.11 & 0.14\\
NGC7331 &  20 & 1.86 &   0.13 &   1.33 &   0.16 &   1.67 &   0.22 &   2.20 &   0.20 & 244.00 &  10.90 &  10.25 & 1.04 \\
\hline
Mean   	&    &  2.07 & 0.187  &   1.64 & 0.211    &      1.87 &  0.213   &      2.48 & 0.205  & 168.83 & 10.45 & 9.93 & 0.72     \\
                                                                              
\enddata
\tablenotetext{a}{Four forms of stability parameters ($\log Q$) are calculated using a fixed $\sigma_g = 11km/s$ and then fitted using a constant. Only the intermediate disk data is used during this fitting. The fitted value and the corresponding  deviation value, $\epsilon$, for each galaxy are tabulated here.  The maximum velocity $V_m$, total stellar mass $M_S$, total gas mass $M_g$ and the ratio between the integrated central disk SFR and intermediate disk SFR $\rm \frac {SFR_{cen}}{SFR_{int}}$  are also tabulated here (data from L08). The galaxies are listed in ascending order of total stellar mass $\log(M_s)$.}
\end{deluxetable}

\begin{figure}
\begin{center}
\includegraphics[scale=0.8]{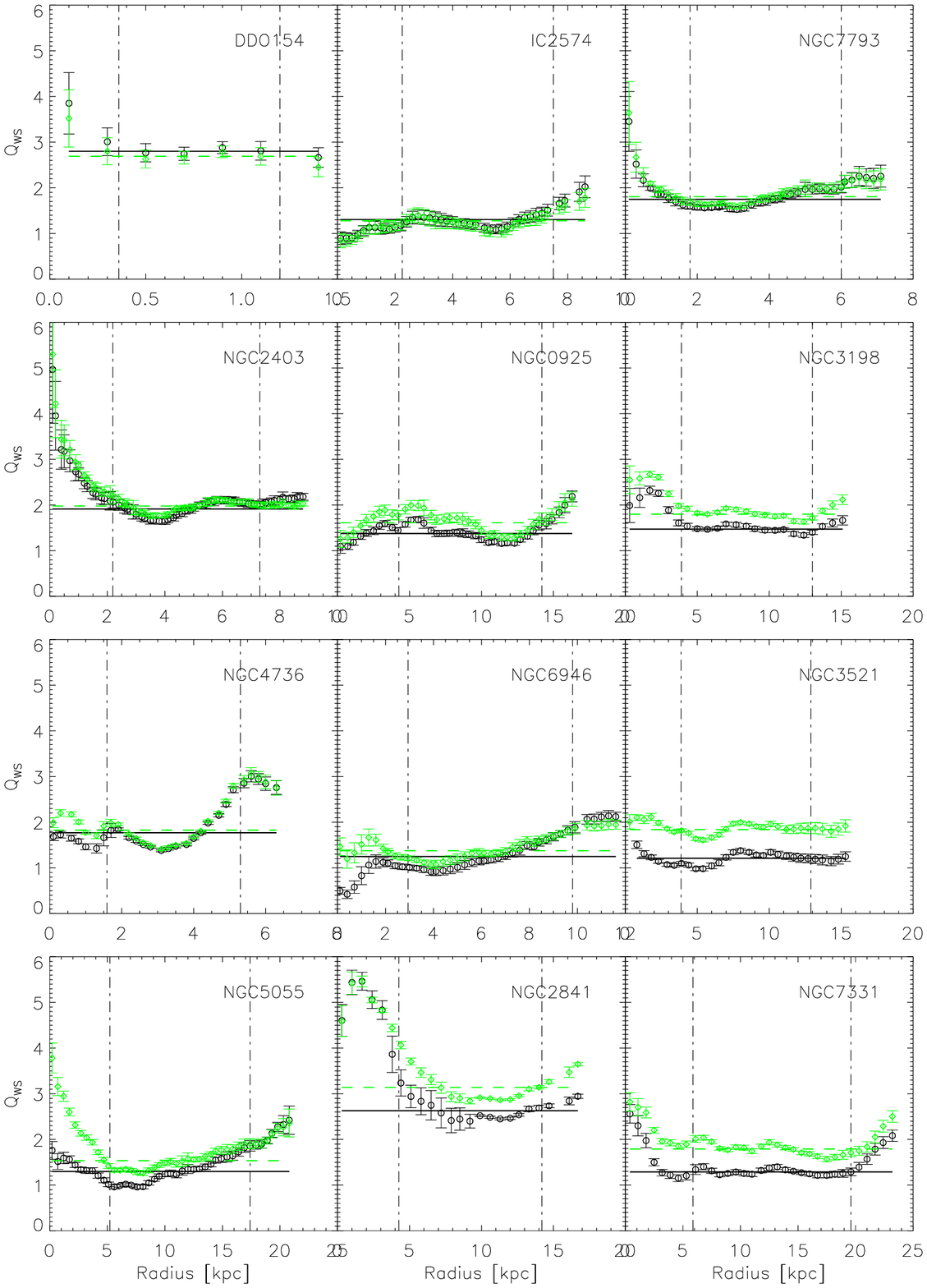}  \caption{$Q_{\rm WS}$ radial profiles.  Black open circles  are  $Q_{\rm WS}$ calculated using $\sigma_g=11km/s$ and black solid lines are their best constant fit; green open diamonds are $Q_{\rm WS}$ calculated using measured $\sigma_g$, and green dash lines are their best  constant fit. Error bars show 1$\sigma$ uncertainty.}
\label{q_vsig}
\end{center}
\end{figure}

\begin{figure}[htbp]
\begin{center}
\includegraphics[scale=0.7]{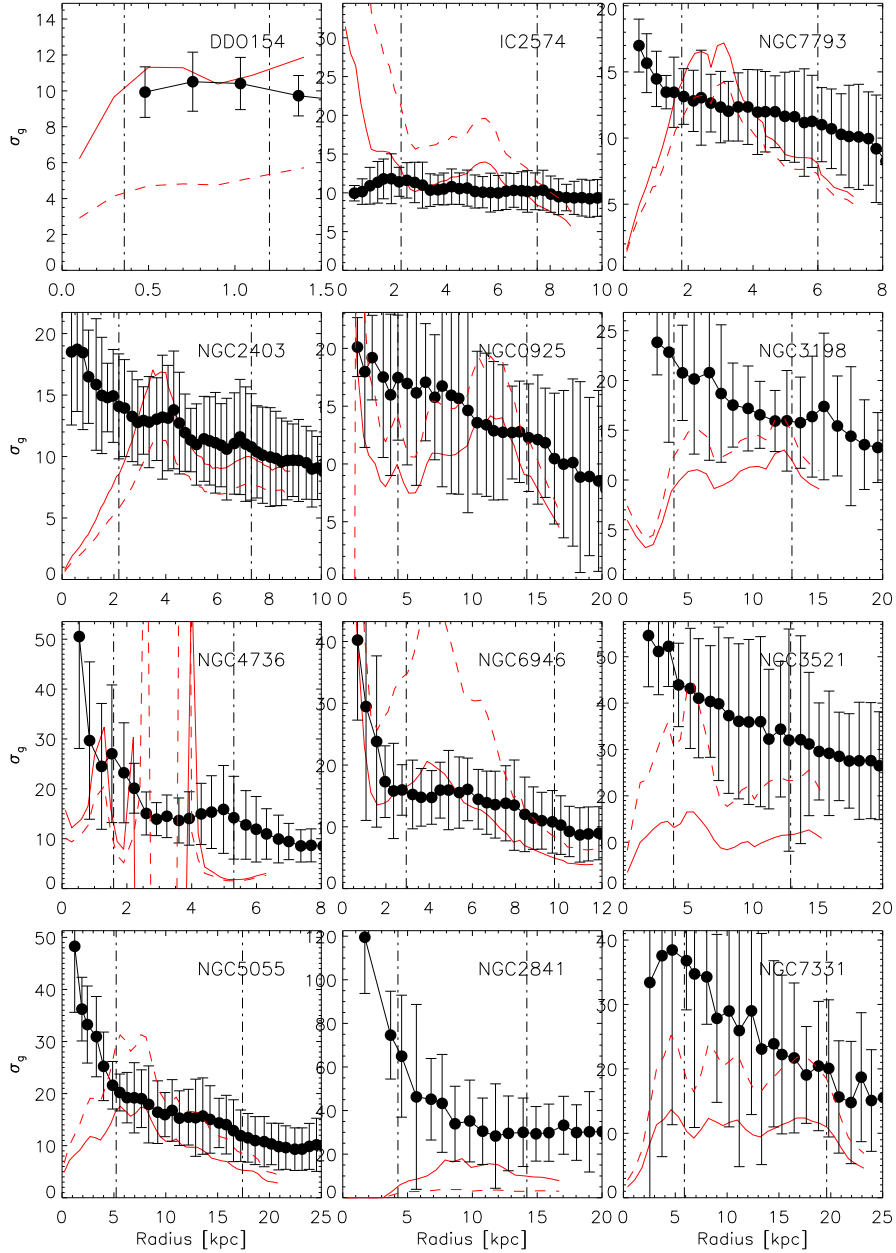}  \caption{Gas velocity dispersion $\sigma_g$  needed in order to keep a constant $Q_{\rm WS}$ with observed rotation curve and $\Sigma_g$ and $\Sigma_s$. Red solid and dashed lines are the needed gas velocity dispersion calculated using the average $Q_{\rm 2f}$ value for the particular galaxy from Table \ref{qerr} and a fixed $Q_{\rm WS}=1.65$ respectively. 
Measured gas velocity dispersion with 1$\sigma$ uncertainty are also over-plotted as black dots and error bars. The $\sigma_g$ data are from Primo \&\ Walter (private communication). }
\label{fig_vsig}
\end{center}
\end{figure}

Figure \ref{q_vsig} shows $Q_{\rm WS}$ for each galaxy using two cases for the gas velocity dispersion: a fixed $\sigma_g$(=11km/s; L08)  and the measured $\sigma_g$. Apparently, there is little difference in these two situations. Furthermore, we calculate the $\sigma_g$ needed to maintain a constant $Q_{\rm WS}$ and plot it on top of the measured $\sigma_g$ in Figure \ref{fig_vsig}. In most cases or at nearly all radii, the needed $\sigma_g$ is within 1$\sigma$ uncertainty of the data and very close to a constant $\sigma_g=11km/s$ as well. Therefore, in order to keep the simplicity of our model, we use the fixed $\sigma_g (=11km/s)$ in the following calculations.

\subsection{Neutral and molecular hydrogen content}
\label{result_gas}
\begin{figure}
\begin{center}
\includegraphics[scale=0.8]{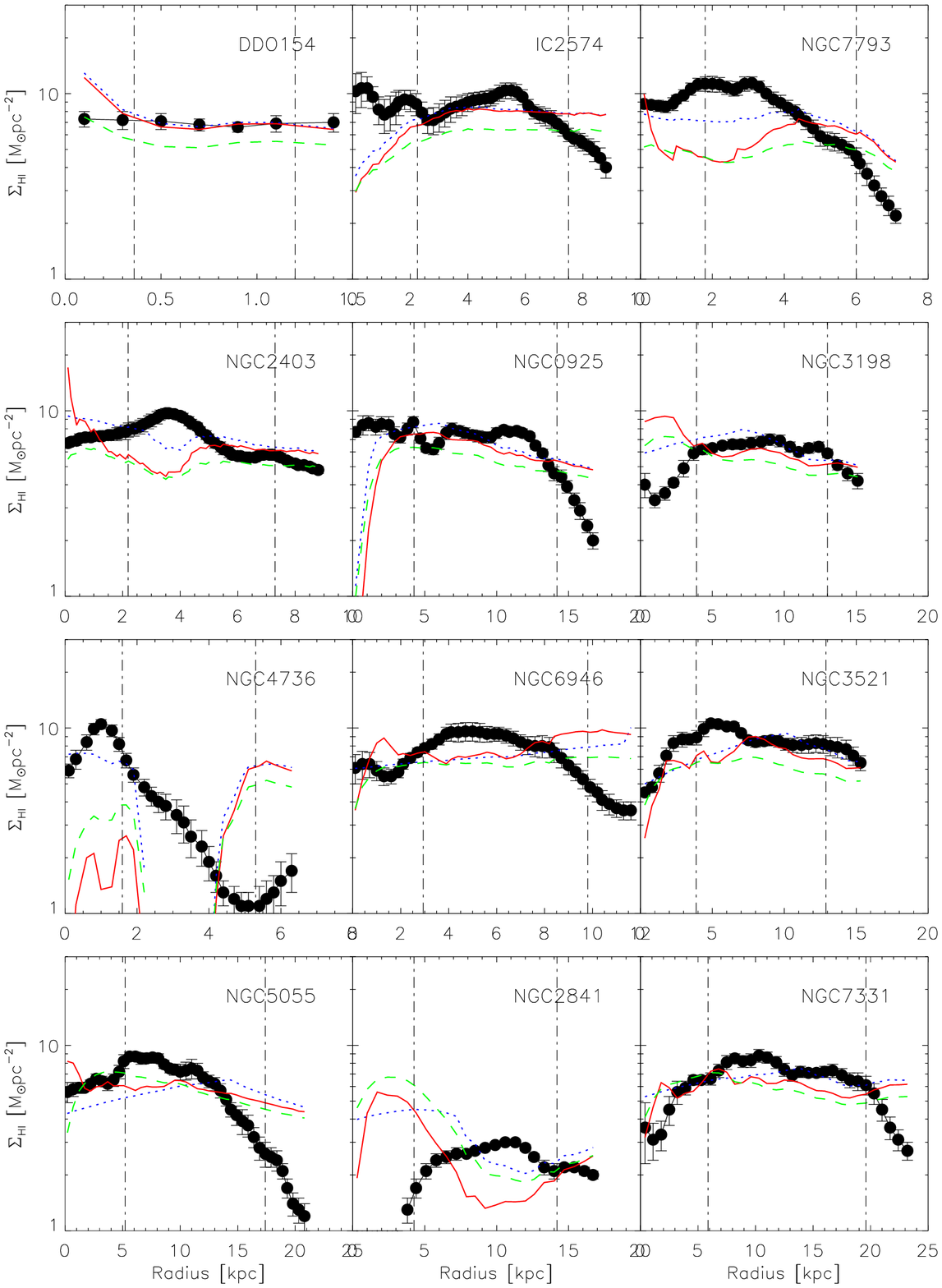}   \caption{The $\Sigma_{\rm HI}$ radial profiles. Black dots and error bars are measured data with 1$\sigma$ uncertainty; and red solid, green dash and blue dot lines are model derived $\Sigma_{\rm HI}$ using SR, PR and KR $R_{\rm mol}$ relations (cf. section \ref{rmol}) respectively. }
\label{Sigma_HI}
\end{center}
\end{figure}

\begin{figure}
\begin{center}
\includegraphics[scale=0.8]{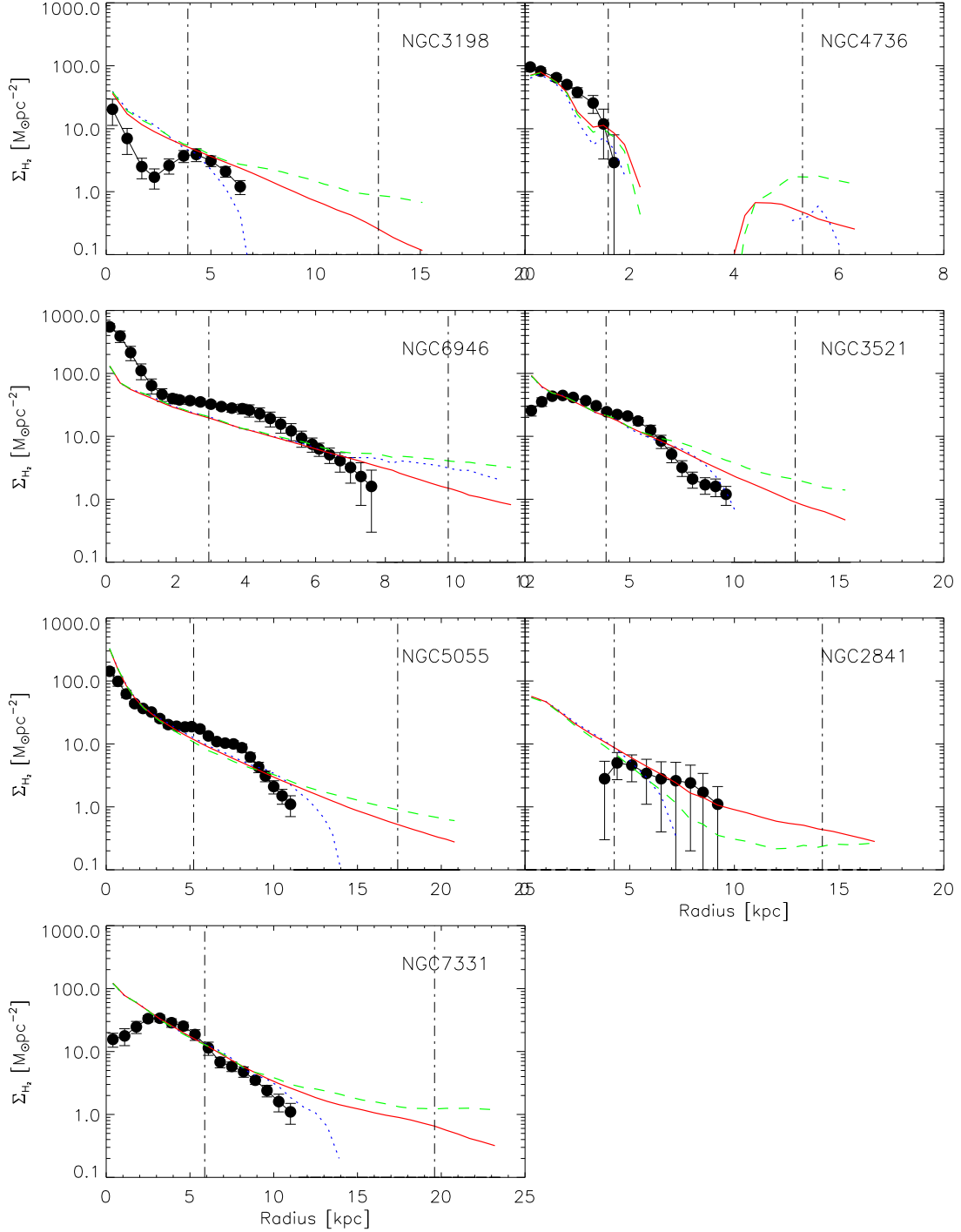}  \caption{The $\Sigma_{H2}$ radial profiles. Conventions follow Fig.\ref{Sigma_HI}.  There are only 7 galaxies shown here because we do not have CO data for the other 5 galaxies. }
\label{Sigma_H2}
\end{center}
\end{figure}

We next calculate the neutral and molecular gas surface densities using the three $R_{\rm mol}$ relations: SR, PR and KR (cf. section \ref{rmol}). We use the fitted $Q_{\rm WS}$ values tabulated for each galaxy in Table \ref{qerr} to derive the gas distribution.   The calculated $\Sigma_{\rm HI}$ and $\Sigma_{\rm H_2}$ radial profiles are shown in Fig. \ref{Sigma_HI} and Fig. \ref{Sigma_H2} respectively. 
We calculate deviation values $\epsilon_{\rm HI}$ and $\epsilon_{\rm H2}$ for the three $R_{\rm mol}$ relations ($\epsilon_{\rm HI,SR}$, $\epsilon_{\rm HI,PR}$, etc.) using equations analogous to Eq. \ref{eq_epsilon_Q} to compare how well the $\Sigma_{\rm HI}$, and the $\Sigma_{\rm H_2}$ observations and model agree. A cursory comparison of Fig. \ref{Sigma_HI} with Fig. \ref{Sigma_H2} suggests that the HI profiles do not fit the data as well as the H$_2$ profiles.  However, this is largely illusory - the $\Sigma_{\rm HI}$ profiles are much flatter than the $\Sigma_{\rm H_2}$ profiles for which a larger display range is required.  Comparison of the $\epsilon_{\rm HI}$ and $\epsilon_{\rm H_2}$ values in Table 2 demonstrate that in general the HI profiles are fit by the model about as well or better than the H$_2$ profiles.
All three $R_{\rm mol}$ model predictions agree with the data (cf. Table \ref{rmol_tab}) reasonably well with the average $\epsilon_{\rm HI} \leq 0.19$ and the average $\epsilon_{\rm H2} \leq 0.29$, equivalent to fractional errors better than 0.6 and 0.9 respectively. The KR relation is the best  in terms of the $\epsilon_{\rm HI}$ values  and the SR relation is the best in terms of $\epsilon_{\rm H_2}$ values.
However, in some low metallicity regions and/or galaxies (e.g. DDO 154 and IC 2547) the KR relation results in no molecular gas. 
We do sometimes have a very bad fit (e.g. $\Sigma_{\rm HI}$ of NGC 2841) or even failure (e.g. $\Sigma_{\rm HI}$ of NGC 4736). The failure is caused by the big dip in the measured $Q_{\rm WS}$ curve (Fig. \ref{q_vsig}) which means the disk is already nearly unstable from just the stars.

The predicted HI surface mass density is usually flat over the radii
considered here, but the model sometimes has a big divergence from the data in the outer regions 
where the observed HI has a steep decline, e.g. in NGC 925 and NGC 6946.

Note that the $  R_{\rm mol}$ relations are tested using our CQ-disk model predicted $  \Sigma_{\rm g}$, not the observed total gas surface mass density. The best $  R_{\rm mol}$ relation we find here need not necessarily be the best one would find when applying to observed ISM profiles.  However, our results are consistent with those of L08, who found the SR relation best in determining the molecular contents using observed values.

\begin{deluxetable}{l c c c c c c c c c c c}

  \tabletypesize{\footnotesize} \tablewidth{0pt} \tablecolumns{12}
  \tablecaption{Model-observation deviation for gas surface densities }
\tablehead{ \colhead{Galaxy}  & \colhead{$N_{\rm ann}$} & \colhead{$\epsilon_{\rm HI_{\rm SR}}$} & \colhead{$\epsilon_{\rm HI_{\rm PR}}$} & \colhead{  $\epsilon_{\rm HI_{\rm KR}}$}    }
\startdata
    \label{rmol_tab}

DDO154 &            4 & 0.02 & 0.13 & 0.02 \\
IC2574 &           27 & 0.06 & 0.14 & 0.06 \\
NGC7793 &           22 & 0.24 & 0.27 & 0.13 \\
NGC2403 &           33 & 0.18 & 0.21 & 0.10 \\
NGC0925 &           22 & 0.08 & 0.13 & 0.08 \\
NGC3198 &           14 & 0.06 & 0.11 & 0.05 \\
NGC4736 &            9 & 0.64 & 0.51 & 0.53 \\
NGC6946 &           24 & 0.13 & 0.13 & 0.11 \\
NGC3521 &           18 & 0.10 & 0.16 & 0.09 \\
NGC5055 &           25 & 0.14 & 0.11 & 0.17 \\
NGC2841 &           15 & 0.25 & 0.23 & 0.20 \\
NGC7331 &           20 & 0.10 & 0.13 & 0.06 \\

\hline
Median &  & 0.13 & 0.14 & 0.10 \\
Mean &   & 0.17 & 0.19 & 0.14 \\

\hline
\hline
& & \colhead{$\epsilon_{\rm H2_{\rm SR}}$} & \colhead{$\epsilon_{\rm H2_{\rm PR}}$} & \colhead{  $\epsilon_{\rm H2_{\rm KR}}$}  \\
\hline

NGC3198 &            4 & 0.20 & 0.24 & 0.31 \\
NGC6946 &           17 & 0.23 & 0.26 & 0.24 \\
NGC3521 &           12 & 0.24 & 0.35 & 0.21 \\
NGC5055 &           12 & 0.20 & 0.25 & 0.20 \\
NGC2841 &            8 & 0.13 & 0.38 & 0.47 \\
NGC7331 &            8 & 0.21 & 0.25 & 0.17 \\
\hline
Median & & 0.21 & 0.26 & 0.24 \\
Mean & & 0.20 & 0.29 & 0.27 

\enddata
\tablenotetext{a}{ $\epsilon$ is calculated  for all three $R_{\rm mol}$ prescriptions. 
The upper half of this table is for $\Sigma_{\rm HI}$ and the lower half is for $\Sigma_{\rm H_2}$.   
There are less  $N_{ann}$ values in the lower half because there are fewer data points for $\Sigma_{H_2}$.
} 
\end{deluxetable}

\subsection{Star formation rate}\label{result_sfr}
\begin{figure}
\begin{center}
\includegraphics[scale=0.7]{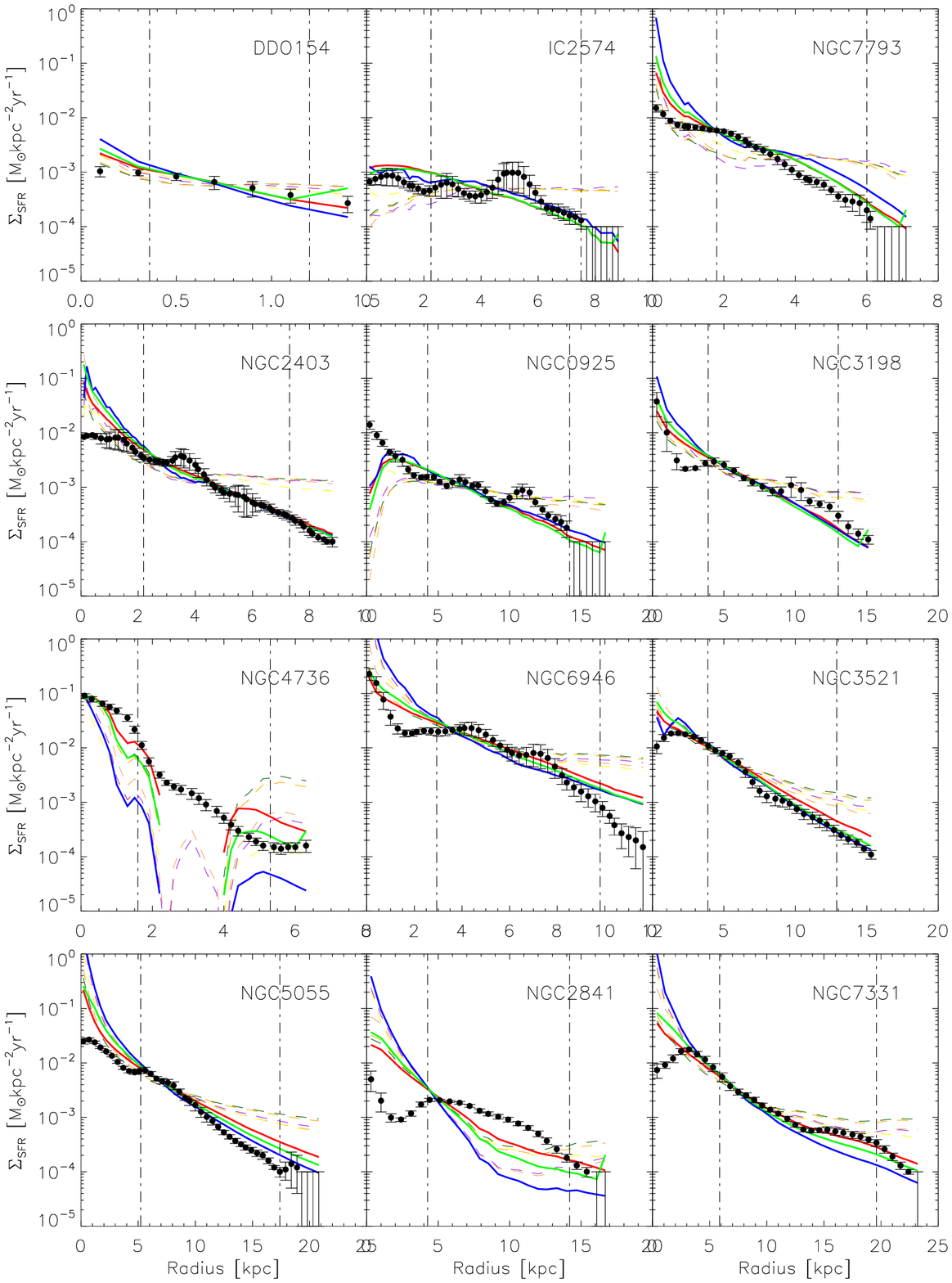}  \caption{$\Sigma_{\rm SFR}$ radial profiles using the SR prescription for $R_{\rm mol}$. Black dots and error bars are measured $\Sigma_{\rm SFR}$ from L08. Thick solid lines are the three best fit SFLs: red, linear molecular SFL; blue, turbulence regulated SFL; green, two-component SFL.  Dash lines are other  SFL predictions: dark green, SK law; orange, free-fall with fixed scale height; sandy brown, free-fall with variable scale height; yellow, orbital time scale; purple, GMC collision.}
\label{SFR_s}
\end{center}
\end{figure}

\begin{figure}
\begin{center}
\includegraphics[scale=0.8]{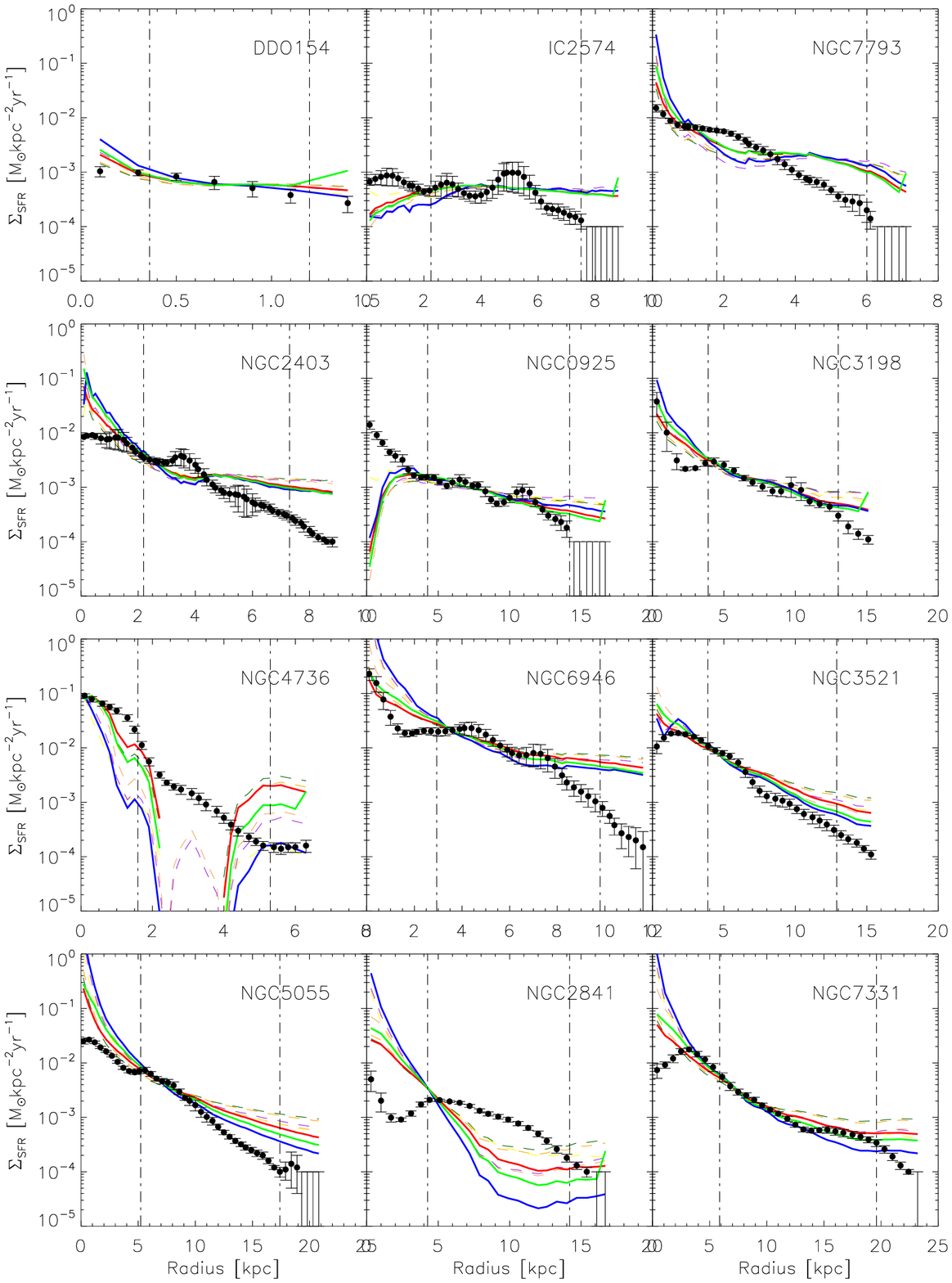}  \caption{$\Sigma_{\rm SFR}$ radial profiles using the PR prescription for $R_{\rm mol}$. Conventions follow Fig.\ref{SFR_s}. }
\label{SFR_p}
\end{center}
\end{figure}

\begin{figure}
\begin{center}
\includegraphics[scale=0.8]{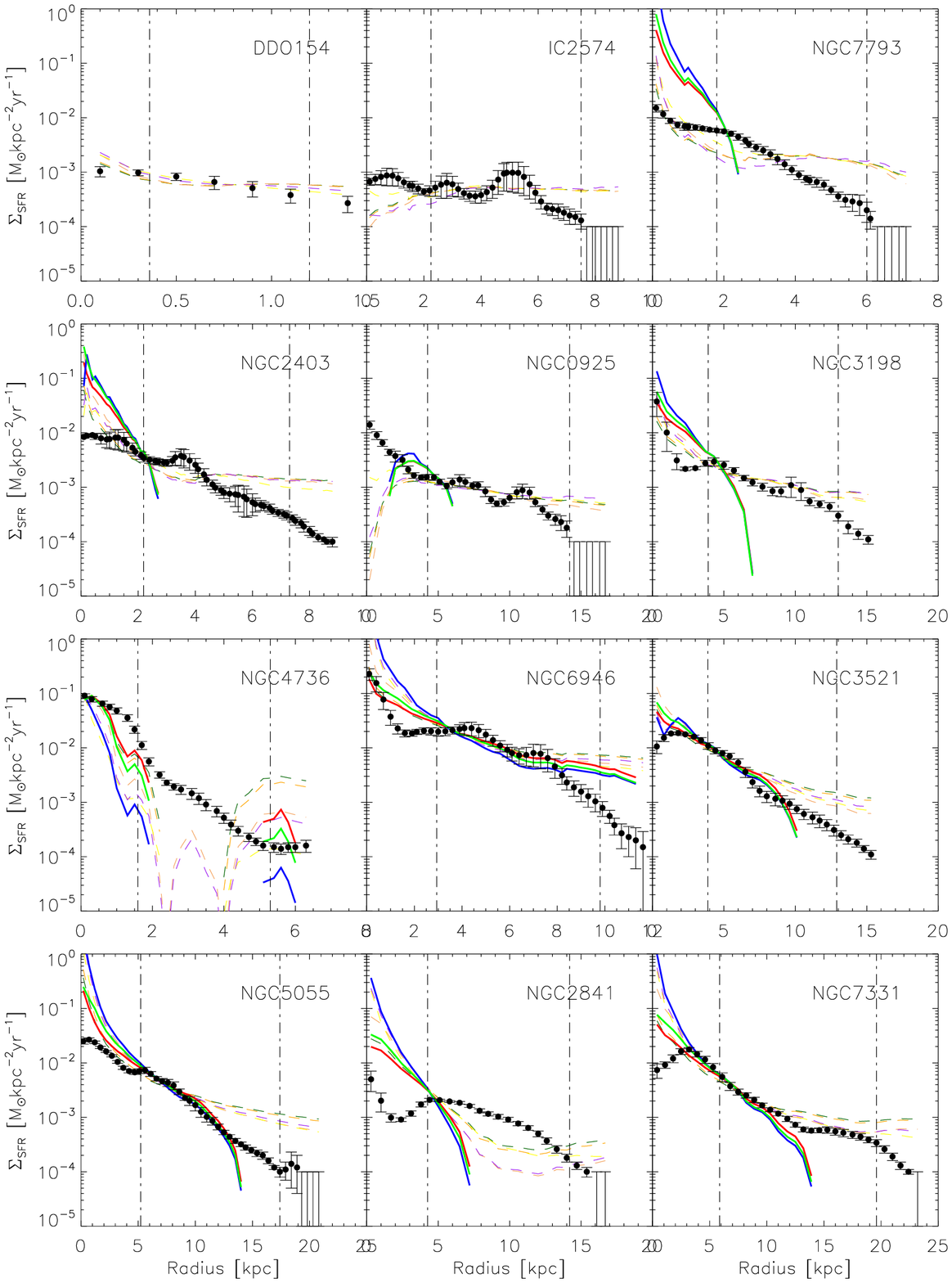}  \caption{$\Sigma_{\rm SFR}$ radial profiles using the KR prescription for $R_{\rm mol}$. Conventions follow Fig.\ref{SFR_s}.}
\label{SFR_k}
\end{center}
\end{figure}

Following \citet{tan10}, we test all SFLs introduced in section \ref{SFLs} using the derived $\Sigma_{\rm HI}$ and $\Sigma_{\rm H_2}$ (cf. section \ref{result_gas}) as the input gas surface mass density values.  For each galaxy, we derive the best-fit coefficients (e.g. $A_{\rm SK}$, $B_{\Omega}$, etc.) for these SFLs and measure the resultant uncertainty, $\epsilon_{SFR}$, which is defined in a similar way to Eq. \ref{eq_epsilon_Q}.  
Since we have three different $R_{\rm mol}$ models, we also have three different sets of $\Sigma_{\rm HI}$ and $\Sigma_{\rm H_2}$ (cf. section \ref{result_gas}) and therefore three different fitted SFL coefficients.The coefficients and the corresponding $\epsilon$ values are listed in Table \ref{sfl_parameter1} and \ref{sfl_parameter2}. The resulting $\Sigma_{\rm SFR}$ profiles from the SFLs are plotted on top of the observed data in Figs. \ref{SFR_s}, \ref{SFR_p}, and \ref{SFR_k}.

Comparing the curves in Fig. \ref{SFR_s}, \ref{SFR_p}, and \ref{SFR_k} 
the SFLs that depend on separating the molecular and neutral phases generally fit better than the SFLs that just depend on the total cool and cold ISM content.  The predictions based on the SR relation performs better than PR and KR relations in terms of SFR predictions.  The KR relation generally performs the worse in terms of SFR predictions.  In addition, we see from Table~\ref{sfl_parameter2} that the coefficient of the fit strongly correlates with stellar mass (as seen in the ordering of the table rows) when using the KR prescription for $R_{\rm mol}$.  This means that SFL with this prescription is not universal.
The $\epsilon_{\rm SFL}$  values in Tables \ref{sfl_parameter1} and \ref{sfl_parameter2} also support the view that the the molecular SFLs using the SR $R_{\rm mol}$ relation generally work the best. 
The exception to this rule is the dynamical time SFL (SFR$_\Omega$) which outperforms the turbulence regulated molecular SFL (SFR$_{\rm KM}$) using the SR relation and all molecular SFLs using the PR and KR relation.  
However, the values in Tables \ref{sfl_parameter1} and \ref{sfl_parameter2} are the fit results using the intermediate radii data only. 
In terms of the simultaneous fit to all data points it is the third best SFL ($\epsilon_\Omega = 0.39$) following the linear molecular SFL (best, $\epsilon_{\rm H2} = 0.30$) and the two component SFL (second best, $\epsilon_{\rm cc} = 0.34$).  This is slightly different from the results of \citet{tan10} who found the two component   (KMT) SFL to be the best SFL in terms of rms dispersion. The best fit to all data for the linear molecular SFL yields a coefficient $A_{\rm H_2}=9.51\times  10^{-2}\, {\rm M_{\odot}\, yr^{-1}\, kpc^{-2}}$ using the SR relation.  This is slightly higher than the L08 value, $(5.25\pm 2.5)\times 10^{-2}\, {\rm M_{\odot}\, yr^{-1}\, kpc^{-2}}$, but  well within its 2$\sigma$ uncertainty.  The higher coefficient implies 80\%\ more efficient star formation and shorter molecular gas cycling times than those derived by L08.

Note that the GMC collision SFL is meant to be applied to molecular gas rich regions only (cf. section \ref{sfl_cc}). 
It is therefore not surprising that our application of this SFL in regions that are not dominated by molecular gas, 
such as central region of NGC 2841, results in a poor match to the observed $  \Sigma_{\rm SFR}$.
Once more, one should bear in mind that the best SFL is selected based on our model predicted gas contents ($  \Sigma_g$ or $  \Sigma_{\rm H_2}$). This is different from the standard approach of using the  observed $  \Sigma_g$ or $  \Sigma_{\rm H_2}$ to predict $  \Sigma_{\rm SFR}$. 
Most of the best-fit SFL coefficients (A or B coefficients in Table \ref{sfl_parameter1} and \ref{sfl_parameter2}), except the coefficient of the GMC collision SFL ($B_{\rm CC}$) and the coefficient of the KMT SFL ($A_{\rm KMT}$),  are on the order of but a little bit larger than those of \citet{tan10}, who apply the SFLs onto observed gas surface mass densities. 
Since we have slightly different sample galaxies with \citet{tan10}, it is possible to have very different fitting values, however in fact, we do have similar $B_{\rm CC}$ and $A_{\rm KMT}$ values to those of \citet{tan10} for galaxies in the overlapping part of our  and Tan's sample.
The model-observation deviation ($\epsilon$ values in Table \ref{sfl_parameter1} \& \ref{sfl_parameter2}) have similar situation: they are also generally larger than but on the order of those in \citet[][the $\chi$ values]{tan10}. 
For example, for the galaxy NGC 7331, we find  $\epsilon_{\rm SFR}= 0.056$ for the linear molecular SFL
while \citet{tan10} determines  $\epsilon_{\rm SFR}= 0.0493$ for this SFL.
However, the combination of the SR prescription for $R_{\rm mol}$ and the linear molecular SFL is claimed to provide the best results by L08, which is consistent with our results.

\begin{deluxetable}{l c c c c c c c c c c c}

  \tabletypesize{\footnotesize} \tablewidth{0pt} \tablecolumns{12}
  \tablecaption{Star Formation Law Parameters for Sample Galaxies (1)}
\tablehead{ \colhead{Galaxy}  & \colhead{$N_{\rm ann}$} & \colhead{$A_{\rm SK}$} & \colhead{$\epsilon_{\rm SFR_{\rm SK}}$} & \colhead{ $A_{\rm ff}$} 
  & \colhead{  $\epsilon_{\rm SFR_{\rm ff}}$}   & \colhead{  $A_{\rm ff^{\prime}}$}   & \colhead{  $\epsilon_{\rm SFR_{\rm ff^{\prime}}}$} & \colhead{  $B_{\Omega}$}  
  & $\epsilon_{\rm SFR_{\Omega}} $  & $B_{\rm CC}$  & $\epsilon_{\rm SFR_{\rm CC}} $ \\
     & &$(10^{-2})$  &  & $(10^{-2})$  &   & $(10^{-2})$  &   & $(10^{-3})$  &  & $(10^{-3})$     &   \\ 
   }
\startdata
    \label{sfl_parameter1}

DDO154 &  4. &  2.51  &  0.155  &  3.28  &  0.156  &  12.49  &  0.159  &  4.01  &  0.062  &  41.62  &  0.128  \\
IC2574 &  27. &  1.66  &  0.274  &  2.14  &  0.274  &  7.49  &  0.272  &  9.49  &  0.273  &  33.78  &  0.343  \\
NGC7793 &  22. &  6.84  &  0.491  &  8.86  &  0.489  &  32.29  &  0.484  &  11.61  &  0.458  &  42.52  &  0.577  \\
NGC2403 &  33. &  6.71  &  0.394  &  8.72  &  0.393  &  32.04  &  0.387  &  10.98  &  0.308  &  57.46  &  0.407  \\
NGC0925 &  22. &  3.30  &  0.200  &  4.27  &  0.194  &  15.42  &  0.167  &  20.39  &  0.204  &  64.03  &  0.273  \\
NGC3198 &  14. &  4.96  &  0.200  &  6.41  &  0.192  &  22.42  &  0.145  &  10.51  &  0.152  &  40.55  &  0.246  \\
NGC4736 &  9. &  12.38  &  0.838  &  12.65  &  0.856  &  14.13  &  1.394  &  0.63  &  1.077  &  14.77  &  1.293  \\
NGC6946 &  24. &  16.23  &  0.355  &  19.27  &  0.344  &  43.65  &  0.294  &  23.42  &  0.270  &  82.37  &  0.360  \\
NGC3521 &  18. &  5.43  &  0.401  &  6.43  &  0.379  &  14.31  &  0.257  &  6.91  &  0.296  &  20.21  &  0.399  \\
NGC5055 &  24. &  6.83  &  0.520  &  8.58  &  0.505  &  25.73  &  0.418  &  11.27  &  0.418  &  30.35  &  0.468  \\
NGC2841 &  15. &  5.79  &  0.341  &  7.18  &  0.390  &  20.22  &  0.676  &  3.50  &  0.375  &  26.96  &  0.605  \\
NGC7331 &  20. &  4.31  &  0.232  &  5.32  &  0.212  &  14.44  &  0.103  &  5.92  &  0.101  &  14.86  &  0.111  \\

\hline
All & 232 & 9.04 &  0.557 &  11.10 &  0.543 &  28.06 &  0.523 &  12.10 &  0.385 &  39.18 &  0.501 \\
Median & &      5.79 & &       7.18 & &      20.22 & &       10.51 & &       40.55 \\
Mean & &      6.41 & &       7.76 & &      21.22 & &       9.89 & &       39.12 \\

\enddata
\tablenotetext{a}{The fitted SFL coefficients and model-observation deviations: SK, Schimidt-Kennicutt Law; ff, free-fall time scale with fixed scale height; $\rm ff^{\prime}$, free-fall time scale with variable scale height; $\Omega$, orbital time scale; CC, GMC collisions in a shearing disk. These five SFLs only depend on the total gas surface mass density and galactic dynamic parameters. The `All' means the fitting using all valid data points from all the sample galaxies. All the `A' coefficients have units of $M_{\odot}kpc^{-2}yr^{-1}$ and all the `B' coefficients are dimensionless.} 
\end{deluxetable}

\begin{deluxetable}{l c c c c c c  }
  \tablewidth{0pt} \tablecolumns{8}
  \tablecaption{Star Formation Law Parameters for Sample Galaxies (2)}
    \tabletypesize{\tiny} 

\tablehead{
\colhead{Galaxy}  
& $A_{\rm H_2}$ & $\epsilon_{\rm SFR_{\rm H_2}}$ & $A_{\rm KM}$ & $\epsilon_{\rm SFR_{\rm KM}}$ & $A_{\rm KMT}$ & $\epsilon_{\rm SFR_{\rm KMT}}$\\
   &  $(10^{-2})$  &  &   &   & $(10^{-2})$  &  \\

}
\startdata
  \label{sfl_parameter2} 
SR relation\\
\hline

DDO154  &  30.94  &  0.060  &  77.70  &  0.142  &  71.09  &  0.058 \\
IC2574  &  17.70  &  0.255  &  83.59  &  0.204  &  38.85  &  0.251 \\
NGC7793  &  13.22  &  0.169  &  40.33  &  0.291  &  29.35  &  0.168 \\
NGC2403  &  11.33  &  0.108  &  27.99  &  0.165  &  24.20  &  0.125 \\
NGC0925  &  9.83  &  0.195  &  61.09  &  0.165  &  20.74  &  0.232 \\
NGC3198  &  6.90  &  0.181  &  16.08  &  0.193  &  14.19  &  0.229 \\
NGC4736  &  11.49  &  0.459  &  1.19  &  1.654  &  11.71  &  0.651 \\
NGC6946  &  14.80  &  0.159  &  15.28  &  0.197  &  23.46  &  0.146 \\
NGC3521  &  5.02  &  0.167  &  4.08  &  0.090  &  7.83  &  0.095 \\
NGC5055  &  6.72  &  0.269  &  7.81  &  0.135  &  12.51  &  0.196 \\
NGC2841  &  3.75  &  0.296  &  4.96  &  0.827  &  7.25  &  0.485 \\
NGC7331  &  4.37  &  0.056  &  3.84  &  0.261  &  7.65  &  0.138 \\

\hline
All & 9.51 &  0.300 &  8.37 &  0.642 &  15.61 &  0.334 \\
Median &        11.33 & &       16.08 & &      20.74 \\
Mean &        11.34 & &       28.66 & &      22.40 \\

\hline
\hline
PR relation\\
\hline
DDO154  &  3.86  &  0.124  &  11.38  &  0.066  &  8.84  &  0.126 \\
IC2574  &  2.71  &  0.254  &  10.78  &  0.284  &  5.88  &  0.253 \\
NGC7793  &  8.03  &  0.432  &  19.49  &  0.494  &  17.80  &  0.428 \\
NGC2403  &  8.04  &  0.327  &  20.86  &  0.330  &  17.62  &  0.320 \\
NGC0925  &  4.83  &  0.140  &  30.42  &  0.163  &  10.45  &  0.133 \\
NGC3198  &  5.73  &  0.101  &  13.57  &  0.098  &  12.06  &  0.083 \\
NGC4736  &  11.67  &  0.832  &  1.18  &  1.926  &  11.90  &  1.040 \\
NGC6946  &  13.47  &  0.300  &  14.65  &  0.268  &  21.79  &  0.266 \\
NGC3521  &  4.56  &  0.309  &  3.90  &  0.194  &  7.28  &  0.228 \\
NGC5055  &  7.08  &  0.391  &  8.29  &  0.249  &  13.42  &  0.324 \\
NGC2841  &  4.82  &  0.565  &  5.64  &  1.116  &  9.01  &  0.770 \\
NGC7331  &  4.13  &  0.103  &  3.75  &  0.145  &  7.36  &  0.058 \\

\hline
All & 8.66 &  0.440 &  8.13 &  0.503 &  15.08 &  0.407 \\
Median &        5.73 & &       11.38 & &      12.06 \\
Mean &        6.58 & &       11.99 & &      11.95 \\

\hline
\hline
KR relation\\
\hline

NGC7793  &  76.85  &  0.396  &  223.23  &  0.426  &  167.85  &  0.406 \\
NGC2403  &  23.20  &  0.422  &  45.28  &  0.504  &  46.87  &  0.445 \\
NGC0925  &  15.66  &  0.254  &  98.27  &  0.230  &  32.49  &  0.259 \\
NGC3198  &  9.09  &  0.898  &  19.59  &  0.915  &  17.84  &  0.931 \\
NGC4736  &  12.46  &  0.454  &  1.18  &  1.499  &  12.71  &  0.647 \\
NGC6946  &  14.11  &  0.268  &  14.88  &  0.255  &  22.54  &  0.238 \\
NGC3521  &  5.07  &  0.193  &  4.08  &  0.206  &  7.83  &  0.204 \\
NGC5055  &  6.44  &  0.190  &  7.54  &  0.232  &  11.90  &  0.209 \\
NGC2841  &  3.60  &  0.637  &  4.62  &  0.869  &  6.80  &  0.732 \\
NGC7331  &  4.22  &  0.295  &  3.67  &  0.409  &  7.24  &  0.360 \\

\hline
All &  9.38 &  0.441 &  8.18 &  0.639 &  15.10 &  0.475 \\
Median &        12.46 & &       14.88 & &      17.84 & \\
Mean &        17.07 & &       42.23 & &      33.49 & \\

\hline

\enddata
\tablenotetext{a}{The fitted SFL coefficients and model-observation deviations: H$_2$, linear molecular SFL; KM, turbulence-regulated SFL; KMT, two-component SFL. These SFLs depend on the molecular gas surface mass density and galactic dynamic parameters. Therefore we list fitted values for $\Sigma_{\rm H_2}$ calculated using both SR and PR relations. Results for the KR relation in DDO154 and IC2574 are omitted because eq.~\ref{eq_molk} yields non real results over much of these galaxies. All the `A' coefficients have units of $M_{\odot}kpc^{-2}yr^{-1}$.}
\end{deluxetable}

\section{Discussion}
\label{discussion}

\subsection{Variations in $Q_{\rm 2f}$ and $\sigma_g$}
The basic assumption of our model is that the two-fluid stability parameter $Q_{\rm 2f}$ is a constant. As we can see from Fig. \ref{q_vsig},  $Q_{\rm WS}$ is roughly constant, especially at intermediate radii. The variation can be large, up to a factor of 3 over the whole radial range sampled but the variations within the intermediate radii regions are smaller, less than a factor $\sim 1.5$ except for three cases NGC4736, NGC 6946, and NGC5055, where the variations are a factor of $\sim 2$ (see Table \ref{qerr}). There is also systematic variations with galactocentric radius: in 7/12 cases $Q_{\rm WS}$ rises towards the centre, sometimes quite sharply (NGC7793, NGC2403, NGC5055, NGC2841), while in 3/4 of the sample $Q_{\rm WS}$ increases at large radii, mostly beyond $r_{25}$. This systematic variation may cause  over prediction of gas and SFR surface densities in the central and outer regions. A rapid central rise of $Q_{\rm WS}$ can have several explanations, which are discussed in section \ref{centraldisk}. The slow rise of $Q_{\rm WS}$ beyond  $r_{25}$ might be caused by undetected  molecular gas in the outer disk, or may be due to the limited supply of ISM as pointed out by \citet{meu13}.

The gas velocity dispersion is hard to measure \citep{tam09} and has large uncertainties (cf. Fig. \ref{fig_vsig}). However, as shown in Fig. \ref{q_vsig} the shape of the $Q_{\rm WS}$ profiles is not greatly affected by choosing a constant $\sigma_g$ or adopting the observed $\sigma_g$ profiles.  In four cases (NGC3198, NGC3521, NGC2841, and NGC7331) the profiles are noticeably shifted vertically between the two options.  This is because the measured $\sigma_g$ is significantly higher than the assumed value of 11 km s$^{-1}$.
 We can also see from Fig. \ref{fig_vsig} that 
the $\sigma_g$ needed to keep a constant $Q_{\rm WS}$ 
(using the observed $\Sigma_s$ and $\Sigma_g$ in Eq. \ref{Qg}, \ref{Qs} and \ref{Qws})
is pretty constant and close to $11 {\rm km\, s^{-1}}$ and 
well within $2\sigma$ uncertainty of the measured gas velocity dispersion.

It might be surprising that $Q_{\rm WS}$ does not vary much no matter whether we use a constant $\sigma_g= 11 {\rm km s^{-1}}$
or the measured $\sigma_g$. This is because the definition of $Q_{\rm WS}$ (eq. \ref{Qws}) is similar to the 
equivalent resistance of two resistors connected in parallel and  $Q_s$ is usually smaller than  $Q_g$
in the inner and intermediate disk regions. Thus,  $Q_{\rm WS}$ is usually dominated by the $Q_s$ value.
This also illustrates that it is the total stability parameter, $Q_{\rm 2f}$, instead of  $Q_g$ that matters most 
and the interaction between gas and stars  plays an important role in balancing the stability of galactic disk.

\subsection{The central disk}
\label{centraldisk}

Figures \ref{SFR_p} - \ref{SFR_k}, suggest that our model overestimates
$\Sigma_g$ and/or  $\Sigma_{\rm SFR}$ in the central disk for many of the sample galaxies. 
In e.g. NGC 2841, NGC 2403 and NGC 7331, 
the model overestimates  $\Sigma_g$ and $\Sigma_{\rm SFR}$  by more than an order of magnitude. 
This result is similar to that of \citet{qui72} although he used a different stability criterion \citep{gol65}. 
\citet{qui72} argued that  this is because the density-wave-induced shocks, 
which are very strong in the inner parts of the galaxy, 
make the gas in the post shock regions dense enough to be Jeans unstable. 
Thus their stability criteria, which does not consider shocks, 
may not be a good representation for the inner disk stability. 
However, our model does work well in the central part of other galaxies, e.g. NGC 7793, 
and furthermore, our model sometimes underestimates the central disk SFR of galaxies like NGC 6946 and NGC 925. 
Possible reasons for the model-observation discrepancy in the central disk could be 
the  fitted functional form of the rotation curves are not good enough to represent the data (cf. section \ref{rcs}),
or the central disk is not a suitable place to apply our model because it is dominated by a bulge instead of being pure disk.

\begin{figure}
\begin{center}
\includegraphics[angle=90,scale=0.7]{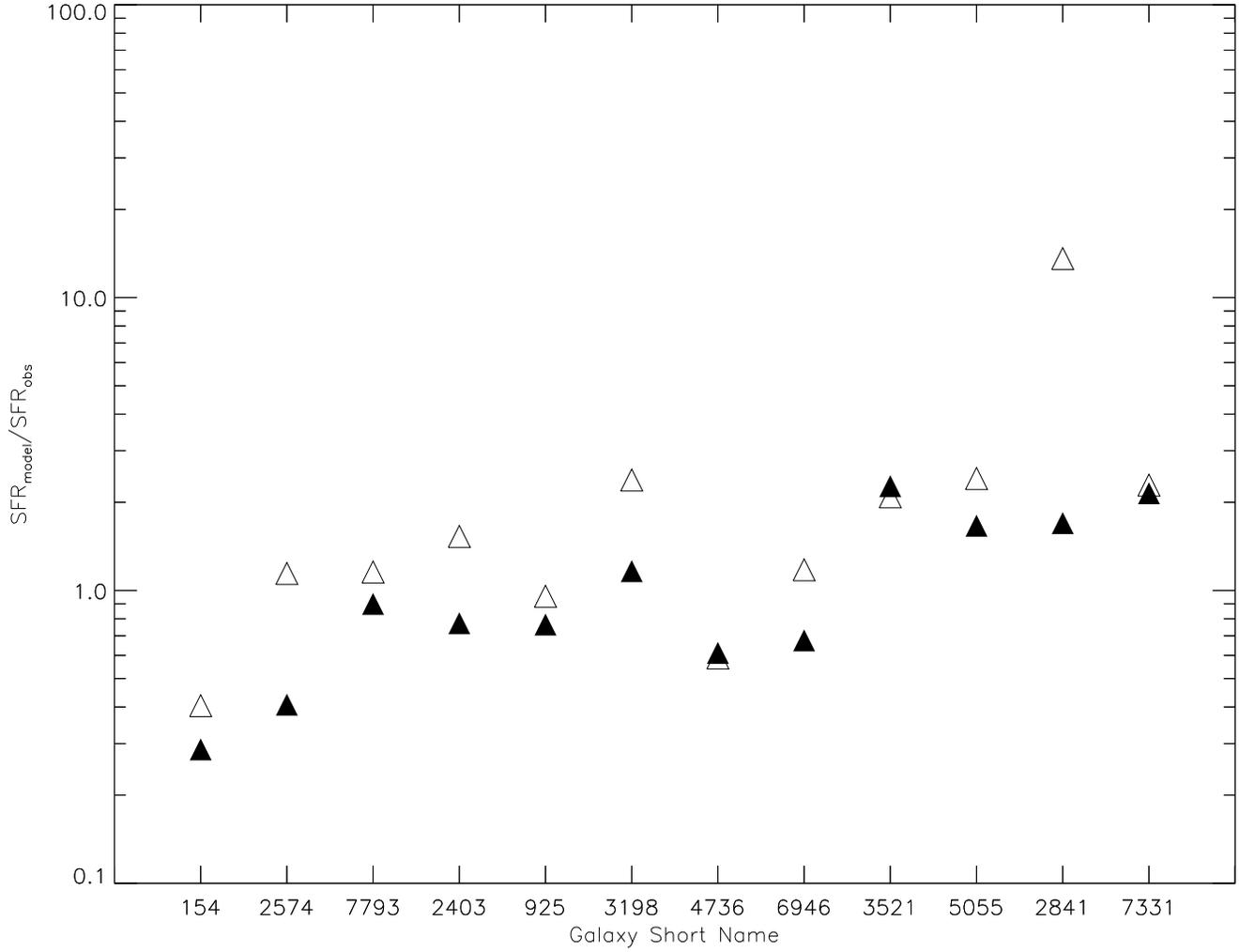} \caption{The ratio of $SFR_{\rm model}/SFR_{\rm obs}$. Solid symbols are for intermediate disk and open symbols are for central disk. The specific model we are using here is the $Q_{\rm WS}$ recipe for $Q_{\rm 2f}$, the SR formulation of $R_{\rm mol}$, and the linear molecular SFL $\Sigma_{\rm SFR,H2}$. 
}
\label{SFR_ratio}
\end{center}
\end{figure}

Despite these complications and the poor fit to the central region in detail,  Fig. \ref{SFR_ratio} shows that the SFR integrated over the central region from our model agrees with the data to a similar level of accuracy as it does over the intermediate radii. The specific model we are using here employs the $Q_{\rm WS}$ recipe for $Q_{\rm 2f}$, the SR formulation of $R_{\rm mol}$, and the linear molecular SFL $\Sigma_{\rm SFR,H2}$. The relatively good agreement in the central region is because usually the model only fails badly in the very central part of the disk,
typically covering an area less than 50\%\  of the  central disk. 
Therefore, the integrated SFR does not deviate from the observation very much.
The exception to this is NGC 2841, where the modelled SFR profiles shown in Figs. \ref{SFR_s} - \ref{SFR_k} deviate strongly from the observed over the entire central disk.

\subsection{Effects of rotation curve parameterisation}
\label{rcs}
\begin{figure}
\begin{center}
\includegraphics[scale=0.7,angle=90]{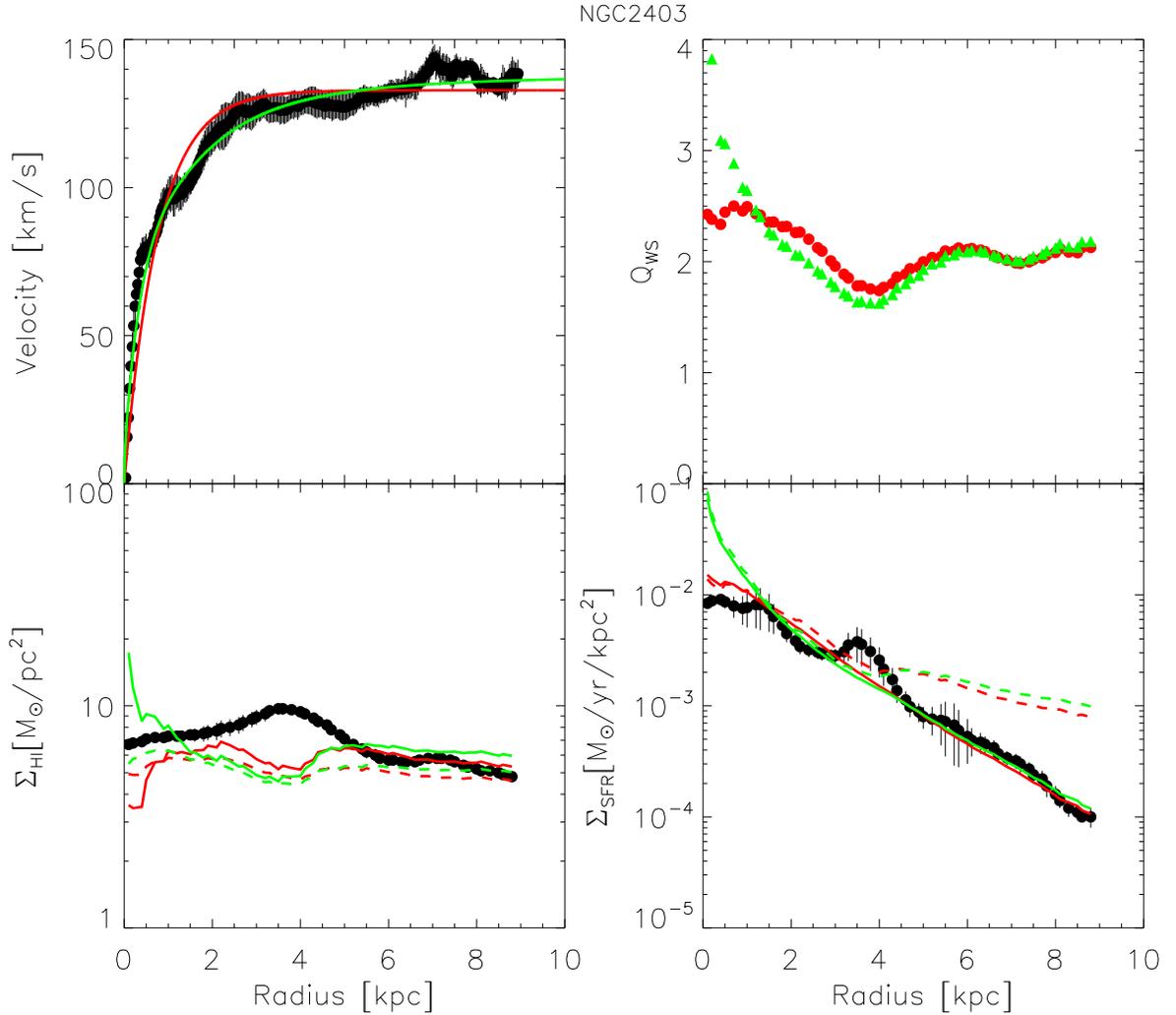} \caption{The effects on the results for NGC 2403 from using different rotation curve fitting functions.
Upper-left panel: Rotation curve. Black dots with error bars are rotation curve data and uncertainties, the red solid line is the URC fit and the green solid line is the exponential rotation curve fit. 
Upper-right panel: Two-fluid stability parameter, $Q_{\rm WS}$. Red dots are calculated using the fitted URC form and green triangles are calculated using fitted exponential rotation curve.
Lower-left panel is HI surface mass density and lower-right panel is SFR surface density.  In these panels the black dots with error bars are measured data with uncertainties.  The red lines are calculated using the fitted URC rotation curve and green lines are calculated using the fitted exponential rotation curve, in both cases the appropriate constant $Q_{\rm WS}$ is adopted.  Solid lines are results using the SR  prescription for $R_{\rm mol}$ and dash lines use the PR  prescription. The SFL used here is the linear molecular SFL.}
\label{NGC2403_exp}
\end{center}
\end{figure}

Since $Q$ is dependent on the derivative of the rotation curve (through $\kappa$), 
the precise shape of the smooth fit must have a quite strong effect on the result, especially in the inner region,
where the rotation curve rises rapidly. In order to test this effect, we use the exponential parameterisation to fit the RC of NGC 2403  and use these new fitting results to recalculate the gas and star formation distribution. The results are shown in Fig.~\ref{NGC2403_exp}. The exponential form results in a shallower slope in the rising part of the RC and therefore a lowered gas and star formation surface density in the central region compared to the result of the URC fitting; whilst the URC form fitting results in a steeper slope and thus a higher gas and star formation surface density in the central region.   As expected, different forms of RC fittings make little difference in the middle and outer disk for galaxies with a flat RC.

\section{Tests and applications of the CQ-disk model} 
\label{application}

Examination of Figs. \ref{Sigma_HI} -- \ref{SFR_k} and Tables
\ref{rmol_tab} - \ref{sfl_parameter2} demonstrate that the CQ-disk model
does a reasonable job (factor of $\sim 2$ agreement) of matching the
overall shape of the \HI, CO, and SFR radial profiles especially at
intermediate radii. Figure \ref{q_vsig} indicates that these radii are
where $Q_{\rm 2f}$ is not only fairly flat but at its minimum, i.e.\ the
most unstable but uniformly so. This may be because at these radii star
formation feedback is effective and the timescales for ISM flows are
sufficient for profiles to evolve to nearly constant stability. As
emphasised in sec 6.2, the model has the most difficulties fitting the
detailed profiles in the central regions of galaxies, but still is
fairly reasonable in predicting quantities integrated out to $r = 0.3
r_{25}$.

Even at intermediate radii, the model curves do not match the profiles
in close detail, with the observed profiles having small scale bumps,
dips, and kinks. Often these local enhancements can also be seen in the
profiles of $Q_{\rm 2f}$, HI, H$_2$, and SFR (e.g.\ the kinks in the
profiles in IC~2574 and NGC~3521 at $R \sim 5.5$ kpc). These indicate
that local enhancements in the gas density are reflected in the SFR
profiles as expected by any reasonable SFL, but result in the $Q$
profiles not being flat in detail. While disks may be evolving towards a
uniform stability, local deviations may build up due to processes beyond
the scope of this model (e.g.\ internal resonances, external
perturbations, minor mergers). Hence the CQ-disk model is not as well
suited to modelling the details of individual galaxies as the overall structure of galaxies in general.

Our sample is relatively small. Further high quality observations would
be useful to further test and optimise our model. New and developing
generations of instrumentation including the JVLA, ASKAP, MeerKAT,
Westerbork+APERTIF for \HI\ observations, ALMA, CARMA, and SMA for CO
(and thus H$_2$), WISE, PanSTARRS, LSST for modelling stellar content,
combined with star formation surveys such as HAGS \citep{jam04}, SINGG
\citep{meu06}, and 11HUGS \citep{ken08} as well as the wide availability
of images of star formation from GALEX and WISE data mean that it will
be possible to do studies similar to ours for well selected samples of
hundreds or thousands of galaxies in the near future.

Examination of Figures \ref{Sigma_HI} -- \ref{SFR_k} demonstrates that
the various $R_{\rm mol}$ and SFL prescriptions diverge from each other
the most in the outer and central regions of galaxies. For the reasons
discussed in Sec.~6.2 the centre is more difficult to model. So we
concentrate on how further study of LSB galaxies and outer disks may
improve our model. The SR and PR recipes diverge from each other there
because the former only depends on the stellar surface mass density, while the
hydrostatic pressure used in the PR prescription becomes more dominated
by gas. The metallicity dependence of the KR relation combined with the
metallicity gradients typically seen in the bright parts of galaxies
\citep{zkh94,kbg03,mou10} will also drive a radial gradient compared to
the SR and PR prescriptions. LSB regions are also important to test
because of indications of low level star formation in extended UV disk
galaxies \citep{thi05,thi07} and \HI\ dominated regions \citep{big10b}.
This low intensity star formation likely signifies IMF variations
\citep{meu09,lee09b,hel10,gun11} which should be incorporated in
future versions of a more comprehensive CQ-disk model.

Comparison of the integrated \HI\ mass and star formation rate may
provide a sensitive test of our model since Figures 4, and 6-8 show
that \HI\ and star formation are largely segregated in galaxies, and
the divergent behaviour of the various model profiles suggests that the
\HI/SFR ratio is likely to be dependent on the $R_{\rm mol}$ and SFL
prescriptions.  We will test this using data from the SINGG and SUNGG
star formation surveys (Zheng et al. 2013, in prep).  

The CQ-disk model may prove useful in galaxy simulations. For example,
currently when generating the initial distribution of particles for N body
simulations of interacting galaxies one often creates disks with the same
scale length for stars and gas \citep[e.g.][]{jnb09,bc11} or giving the
stellar disks exponential profiles with differing scale lengths
\citep[e.g.][]{ljcp08}. Assuming a constant $Q_{\rm 2f}$ disk provides
alternative easy to implement initial conditions for detailed
simulations of galaxies. One could also use the CQ-disk model to enhance
semi-analytic cosmological simulations, i.e.\ in a manner similar to
that done by \citep{duf12a,duf12b} who modeled the detectability of
galaxies in future \HI\ surveys. A similar application of the CQ-disk
model will allow better \HI\ line profile models, as well as models of
H$_2$ (CO), and star formation in the same volume.

\section{Summary}
\label{conclusion}
We have developed a simple `constant $Q$ disk'   (CQ-disk)  model for predicting the distribution of ISM and star
formation in galaxies based on the assumption that the two-fluid instability parameter ($Q_{\rm 2f}$) of the galactic disk is a constant. The model predicts  the gas surface mass density and star formation intensity given the rotation curve, stellar surface mass density and the gas velocity dispersion. 
In this paper we compared radial profiles of HI, and H$_2$ surface mass density and star formation intensity from a sample of 12 galaxies from L08.  In order to optimise our model, we tried various prescriptions for calculating $Q_{\rm 2f}$, the ratio $R_{\rm mol} = \Sigma_{\rm H2}/\Sigma_{\rm HI}$, and the star formation law (SFL).
We find that

\begin{itemize}
\item The $Q_{\rm 2f}$ profiles are fairly flat over the intermediate radii of the disk, with variations of a factor of $\sim 1.6$ about the mean, no matter which recipe of $Q_{\rm 2f}$ is employed.  The \citet{raf01} formulation of $Q_{\rm 2f}$ has the strongest physical basis of the recipes we tried, and also marginally the flattest $Q_{\rm 2f}$ profiles. However it is the most difficult to implement since it requires the wavelength of the most unstable mode to be derived.  The \citet{wan94} approximation, is the most practical recipe for $Q_{\rm 2f}$ in terms of  its very simple form, and the one we have adopted in our model.

\item We tested three prescriptions of $R_{\rm mol}$ by comparing the observed surface mass densities of neutral and molecular gas with our model predictions.  All three prescriptions produce $\Sigma_{\rm HI}$ and $\Sigma_{\rm H_2}$ profiles that match the observed profiles with typical variations better than a factor of 2 about the mean.  Overall, the empirical scaling of  $R_{\rm mol}$ with stellar surface mass density (SR) proposed by \citet{ler08} produces the best fits to both the $\Sigma_{\rm HI}$ and $\Sigma_{\rm H_2}$ profiles, and therefore we favor this model. While the \citet{kru08,kru09a,kru09} self-shielding prescription KR produces typically better matches to the $\Sigma_{\rm HI}$ profiles than the SR relation, the matches to the $\Sigma_{\rm H_2}$ profiles are worse. 
Although the best $  R_{\rm mol}$ relation selected here is based on the predicted $  \Sigma_{\rm g}$ from our CQ-disk model, it is consistent with the L08 results based on observed neutral and molecular ISM profiles.

\item We tested eight versions of the SFL, five that depend on just the total ISM content (defined as combined neutral and molecular component) and three that require the ISM be separated in to neutral and molecular phases. The latter three were tested with all three prescriptions of $R_{\rm mol}$ that we trialled.  The linear molecular SFL, SFR$_{\rm H2}$ produces the best matches to the $\Sigma_{\rm SFR}$ profiles when used with the SR $R_{\rm mol}$ relation, with the models typically agreeing with the observed profiles to within a factor of 2.  The two-component SFL SFR$_{\rm KMT}$, combined with the SR relation does second best with agreement to typically a factor of 2.2, while the orbital time SFL SFR$_\Omega$, at third best in terms of $\epsilon_{\rm SFR}$ value, is the best of the single component ISM SFLs with agreement to within a factor of $2.5$ over the intermediate radii disk.
Again, the best SFLs here are selected based on our model predicted $\Sigma_{\rm g}$ as well as $\Sigma_{\rm H_2}$, not the observed gas surface mass densities. However, the results are consistent with those from L08 and \citet{tan10} who do use observed ISM profiles for their tests. This consistency also supports our CQ-disk model in an indirect way.

\item The modelled star formation intensity profiles (Figs \ref{SFR_s}, \ref{SFR_p}, \ref{SFR_k}) match the observations best at intermediate radii, and show the largest deviations in the central region (where $R <= 0.3 R_{25}$).  
This suggest that a more elaborate model is needed to explain the ISM and gas
distribution in the centres of galaxies, especially those with bulges or a steeply rising rotation
curve. However, integrating the SFR over the entire central region area we find that the models agree with the observations within a factor of 3 in all but one case.  This indicates the systematic discrepancy integrated over  galaxy centres generally is not severe.

\end{itemize}

Since we are testing a model for the distribution of star formation and gas in galaxies that we base on inferences from THINGS team results, there is no surprise that our model works well for large portions of galaxies selected from the THINGS sample. Our future papers will test our model on galaxies selected independently.

One advantage of our models is that they are easy to calculate given the rotation curve and the distribution of stellar mass. This may prove to be a useful advantage in terms of implementation compared to  more detailed models like that of \citet{ost10}. However to handle the central regions, some reasonable modifications may be required, such as requiring gas and SFR to only be at  $r > 0.3r_{25}$ for galaxies with $v_{flat}$ above some fiducial value.

\acknowledgments
We thank A. Leroy and E. de Blok for kindly providing the THINGS data. 
We thank the referee very much for thoughtful comments and suggestions, 
which largely improved the content and clarity of our paper.
We also thank Anahi Caldu Primo and Fabian Walter for  providing the gas velocity dispersion data. 
ZZ also thanks ICRAR for hospitality during his visit to Perth, Western Australia.  
This work is funded in part by GALEX GI grant NNX09AF85G.

\end{document}